\documentclass[aps,pre,reprint,superscriptaddress]{revtex4-2}

\pdfoutput=1

\usepackage{graphicx}
\usepackage{amssymb}
\usepackage{amsmath}
\usepackage{xcolor}
\usepackage{pdfpages}
\usepackage{pgffor}
\usepackage{multirow}
\usepackage{gensymb}
\usepackage{grffile}

\allowdisplaybreaks
\usepackage{placeins} % For \FloatBarrier to restrict floats

\makeatletter
\AtBeginDocument{\let\LS@rot\@undefined}
\makeatother

\begin{document}

\title{Analytical Interaction Potentials for Disks in Two Dimensions}

\author{Binghan Liu}
\affiliation{Department of Physics, Virginia Tech, Blacksburg, Virginia 24061, USA}
\affiliation{Center for Soft Matter and Biological Physics, Virginia Tech, Blacksburg, Virginia 24061, USA}
\affiliation{Macromolecules Innovation Institute, Virginia Tech, Blacksburg, Virginia 24061, USA}
\author{Junwen Wang}
\affiliation{Center for Soft Matter and Biological Physics, Virginia Tech, Blacksburg, Virginia 24061, USA}
\affiliation{Macromolecules Innovation Institute, Virginia Tech, Blacksburg, Virginia 24061, USA}
\affiliation{Department of Mechanical Engineering, Virginia Tech, Blacksburg, Virginia 24061, USA}
\author{Gary S. Grest}
\affiliation{Sandia National Laboratories, Albuquerque, NM 87185, USA}
\author{Shengfeng Cheng}
\email{chengsf@vt.edu}
\affiliation{Department of Physics, Virginia Tech, Blacksburg, Virginia 24061, USA}
\affiliation{Center for Soft Matter and Biological Physics, Virginia Tech, Blacksburg, Virginia 24061, USA}
\affiliation{Macromolecules Innovation Institute, Virginia Tech, Blacksburg, Virginia 24061, USA}
\affiliation{Department of Mechanical Engineering, Virginia Tech, Blacksburg, Virginia 24061, USA}

\date{\today}% It is always \today, today,

\begin{abstract}
Compact analytical forms are derived for the interactions involving thin disks in two dimensions using an integration approach. These include interactions between a disk and a material point, between two disks, and between a disk and a wall. Each object is treated as a continuous medium of materials points interacting by the Lennard-Jones 12-6 potential. By integrating this potential in a pairwise manner, expressions for the potentials and resultant forces between extended objects are obtained. All the results are validated with numerical integrations. The analytical potentials are implemented in LAMMPS and used to simulate two-dimensional suspension of disks with an explicit solvent modeled as a Lennard-Jones liquid. In monodisperse disk suspensions, a disorder-to-order transition of disk packing is observed as the area fraction of disks is increased or as the solvent evaporates. In bidisperse disk suspensions being rapidly dried, stratification is found with the smaller disks enriched at the evaporation front. Such ``small-on-top'' stratification echoes the similar phenomenon occurring in three-dimensional polydisperse colloidal suspensions that undergo fast drying. These potentials can be applied to a wide range of two-dimensional systems involving disk-like objects.
\end{abstract}

\maketitle

\section{Introduction}

As one of the most widely used interatomic potentials, the Lennard-Jones (LJ) potential is often considered an archetype for describing molecular interactions because of its functional simplicity and ability to capture key physical aspects of intermolecular forces.\cite{Lennard-Jones1932} It strikes a reliable balance between accuracy and computational feasibility for simulating simple atomic and molecular systems. The LJ potential is also frequently used to investigate the structure, phase behavior, and dynamics of two-dimensional (2D) systems. For example, Rees-Zimmerman \textit{et al.} recently used the LJ potential as an archetypal model to test numerical methods for unraveling inter-particle potentials in 2D colloidal suspensions.\cite{Rees-Zimmerman2025JCP} The LJ potential is the basis of the 2D Mercedes-Benz water model, where water molecules are represented as LJ disks with explicitly added orientation-dependent interactions to capture hydrogen bonding.\cite{Ogrin2024CPC} Caporusso \textit{et al.} recently used a model based on Brownian disks interacting via the LJ potential, combined with a nonconservative transverse force, to study a chiral fluid in 2D.\cite{Caporusso2024PRL}

Compared with their three-dimensional counterparts, 2D systems exhibit distinct features due to enhanced fluctuations and the critical nature of ordering.\cite{Han2008PRE} One famous example is the Kosterlitz-Thouless-Halperin-Nelson-Young (KTHNY) transition,\cite{Kosterlitz1972, Kosterlitz1973, Halperin1978, Nelson1979, Young1979} which describes the melting transition in 2D crystals as proceeding through two continuous transitions involving an intermediate hexatic phase, rather than via a single discontinuous melting transition. The hexatic phase exhibits a combination of short-range positional order and quasi-long-range orientational order.\cite{Strandburg1988, Kosterlitz2017}

The study of 2D systems of particles interacting via the LJ potential has played a crucial role in deepening our understanding of phase transitions in 2D.\cite{Ryzhov2023JETP, Nishikawa2023PRE} Wierschem and Manousakis conducted large-scale Monte Carlo simulation of 2D LJ solids of up to 102,400 particles and confirmed the two-stage melting scenario predicted by the KTHNY theory.\cite{Wierschem2011PRB} Li and Ciamarra performed molecular dynamics (MD) simulations of attractive squares, pentagons, and hexagons, as well as LJ point particles in 2D and found that attraction tends to promote a discontinuous melting transition without the hexatic phase.\cite{LiYanwei2020PRL} In another work, they simulated particles in 2D interacting by the truncated LJ potential and showed that melting occurs via a two-step process at a high temperature but the hexatic-fluid transition is first-order instead of continuous.\cite{LiYanwei2020PRE} Furthermore, the hexatic phase disappears at a lower temperature well above the liquid-gas critical temperature.\cite{LiYanwei2020PRE} A similar two-step melting transition was confirmed by Khali \textit{et al.}, who performed MD simulation of a 2D system of particles interacting via the repulsive LJ potential.\cite{Khali2021SM} Tsiok \textit{et al.} further systematically studied the role of attraction in the phase diagrams and melting behavior of 2D particle systems interacting via the generalized LJ potential with MD simulations and also found that increasing attraction between particles favors a first-order melting transition.\cite{Tsiok2022JCP} Various melting scenarios of 2D systems were recently reviewed by Ryzhov \textit{et al.},\cite{Ryzhov2023JETP} who analyzed the approaches for determining the type of transition and the associated parameters by computer simulations.

Experimental evidence of various melting schemes in 2D has been found with colloidal systems,\cite{Wei1998PRL, Han2008PRE, Wang2010JCP, Gasser2010ChemPhysChem, Thorneywork2017PRL} dusty plasma,\cite{Vasilieva2021SciRep} driven granular matter,\cite{Olafsen2005PRL, Sun2016SciRep} and atoms forming an adsorbed monolayer on a surface.\cite{Dimon1985PRB, Ternes2010ProgSurfSci} Except the last one, many of these systems involve finite-sized particles. Following the procedure developed for three-dimensional systems, one approach to model extended particles in 2D is to use composite particle models.\cite{Nguyen2019} As the counterpart of a sphere in three dimensions, a disk is represented by a group of point particles distributed inside a circle corresponding to the boundary of the disk. However, composite particle models are usually computationally expensive and subjected to nontrivial constraints to maintain the shape of particles. Another approach for modeling finite-sized disks is to develop analytical potentials suitable for their interactions, assuming that a disk is made of a uniform distribution of point masses. Progress along this line for disks in 2D is the focus of this paper.

The practice of deriving interaction potentials between molecules and extended objects by summing or integrating pairwise interatomic potentials dates back to the 1930s,\cite{Bradley1932, Lennard-Jones1932} shortly after London laid a foundation for the dispersion (i.e., van der Waals) forces between nonpolar atoms and molecules with quantum mechanics.\cite{London1930aEng} Hamaker obtained analytical forms of the van der Waals attraction between two spheres, between two plates, and between a sphere and a plate using the integration approach and provided a general framework based on the assumption that interatomic interactions are pairwise additive.\cite{Hamaker1937} Steele advanced the integration approach by summing LJ interactions over discrete lattice sites and introduced the 10-4-3 potential, which was applied to structured surfaces such as graphite.\cite{Steele1973} The integrated form of the full LJ 12-6 potential for two spheres was later reported by Everaers and Ejtehadi.\cite{Everaers2003} More recently, Wang \text{et al.} reported the analytical forms of the integrated LJ potential between two thin rods,\cite{Wang2025PRE} between a thin rod and a point particle,\cite{Wang2025PRE} and between a thin rod and a sphere\cite{Wang2025EPJE} in arbitrary three-dimensional configurations.

Although the integrated form of the LJ 12-6 potential has been reported for spheres\cite{Everaers2003} and implemented in the LAMMPS simulator as the COLLOID package,\cite{LAMMPS_COLLOID} the corresponding potential for thin disks has not been developed. Here, we employ the integration approach to derive the compact forms of the interaction potentials for disks in 2D. Each disk is treated as a continuous medium uniformly filled with LJ material points. Assuming pairwise additivity, the integrated potential for disks is derived, which is expressed as a function of the center-to-center distance between disks, with their radii as parameters. The potentials in 2D between a disk and a LJ point particle, between a disk and a semi-infinite wall, and between a LJ point particle and a semi-infinite wall are also included. These potentials are implemented in LAMMPS as a user package, which can be used to investigate a wide range of 2D systems including colloidal suspensions of mineral clays like nontronite or laponite and disk-like nanoparticles, discotic liquid crystals, and granular materials. The potentials provide a natural description of hard disks and a new platform for studying structural phase transitions in 2D.

The remainder of the paper is organized as follows. The analytical forms of the integrated potentials between two disks and between a disk and a flat wall based on the LJ potential are presented in Sec.~\ref{sec:disk:lj}, with the derivation process detailed in the Supplementary Material. In Sec.~\ref{sec:disk:md_methods}, the liquid-vapor coexisting densities and the critical point are determined for a LJ liquid in 2D, which serves as a solvent for modeling disk suspensions. The equilibrium structure, disorder-order transition, and drying behavior of monodisperse disk suspensions are reported in Sec.~\ref{sec:disk:res}. This part also includes results for suspensions containing a mixture of bidisperse disks, which are subjected to a rapid drying process. Sec.~\ref{sec:disk:conc} summarizes the findings and provides concluding remarks.

%SFC: only show potentials in the main text; expressions on forces will be included in the SI
\section{Integrated Lennard-Jones Potentials for Disks in Two Dimensions}\label{sec:disk:lj}

%SFC: point mass is more commonly used than mass point

To obtain an analytical potential for the interaction between disks, we start with the Lennard-Jones (LJ) 12-6 potential, which describes the interaction between two point masses.
\begin{equation}
    \label{eq:lj_potential}
    U_\text{LJ}(r) = 4\epsilon \left[\left( \frac{\sigma}{r}\right)^{12} - \left(\frac{\sigma}{r}\right)^6 \right]~.
\end{equation}
Here, $\epsilon$ is the energy scale, $\sigma$ can be considered as the characteristic size of the point mass, and $r$ is the distance between the point masses. $\epsilon$ and $\sigma$, together with the mass of a LJ point particle, $m$, define the LJ units. For example, the time unit is $\tau \equiv \sqrt{m\sigma^2/\epsilon}$.

To facilitate the derivation of the integrated potentials for disks, we rewrite the LJ potential as
\begin{equation}
    \label{eq:lj_potential_two}
    U_\text{LJ}(r) = \frac{A_2}{r^{12}} - \frac{A_1}{r^{6}}~.
\end{equation}
Comparing Eq.~(\ref{eq:lj_potential_two}) with Eq.~(\ref{eq:lj_potential}), $A_2 \equiv 4\epsilon \sigma^{12}$ and $A_1 \equiv 4\epsilon \sigma^{6}$. Reciprocally, $\epsilon = A_1^2/(4A_2)$ and $\sigma = \left(A_2/A_1\right)^{1/6}$. The LJ potential in Eq.~(\ref{eq:lj_potential_two}) will be integrated for various geometries to produce analytical forms of the interaction potentials for disks in two dimensions (2D). The main results are summarized below and the details of the derivation process are provided in the Supplementary Material.

\subsection{Integrated Disk-Point Potential}

We first present the results for the interaction potential between a disk and a LJ point mass in 2D, which is needed to simulate the motion of a disk in an explicit solvent modeled as a liquid consisting of LJ point masses. The disk-point potential can be derived by integrating the LJ potential in Eq.~(\ref{eq:lj_potential_two}) for a disk and a point mass. The attractive component of the integrated disk-point potential in 2D is
\begin{equation}
    \label{eq:int_DP_attr_final}
    U_\text{DP}^\text{A}(r) = -\frac{\pi}{2} A_1 \lambda_a \frac{a^2\left(2 r^2 + a^2 \right)}{\left(r^2-a^2\right)^4}~,
\end{equation}
where $\lambda_a$ is the areal density of LJ point masses making up the disk, $a$ is the radius of the disk, and $r$ is the distance between the center of the disk and the point particle. $\lambda_a$ is assumed to be constant here. The repulsive component of the integrated disk-point potential in 2D is
\begin{equation}
    \label{eq:int_DP_rep_final}
    U_\text{DP}^\text{R}(r) = \frac{\pi}{5} A_2 \lambda_a \frac{a^2 \left(5 r^8 +40 a^2 r^6 +60 a^4 r^4 +20 a^6 r^2 + a^8\right)}{\left(r^2-a^2\right)^{10}}~.
\end{equation}
As expected, both attraction and repulsion are reduced to their forms between two point masses in Eq.~(\ref{eq:lj_potential_two}) in the limit of $a \rightarrow 0$ under the constraint of $\pi a^2 \lambda_a = 1$.

\subsection{Integrated Disk-Disk Potential}

The attraction between a disk and a point in Eq.~(\ref{eq:int_DP_attr_final}) can be integrated further for a second disk to obtain the attraction between two disks in 2D. The details are available in the Supplementary Material. The most general form of the integrated attraction for two disks with radii of $a$ and $b$, respectively, is
\begin{widetext}
    \begin{equation}
        \label{eq:int_DD_attr_final}
        U_\text{DD}^\text{A}(r) = -\frac{\pi^2 A_1 \lambda_a \lambda_b}{2} \frac{a^2 b^2 \left[ 2r^4 - r^2 \left(a^2+b^2\right) - \left(a^2-b^2\right)^2 \right]}{\left[(r-a-b) (r-a+b) (r+a-b) (r+a+b)\right]^{5/2}}~.
    \end{equation} 
\end{widetext}
Here, $\lambda_b$ is the areal density of LJ point masses on the second disk and $r$ is the center-to-center distance between the two disks. Similarly, the disk-point repulsion in Eq.~(\ref{eq:int_DP_rep_final}) can be integrated for the second disk to obtain the repulsion between two disks in 2D, the final form of which is
\begin{widetext}
\begin{eqnarray}
    \label{eq:int_DD_rep_final}
    U_\text{DD}^\text{R}(r) &=& \frac{\pi^2 A_2  \lambda_a \lambda_b}{5} \frac{a^2 b^2}{\left[\left(r+a+b\right)\left(r+a-b\right)\left(r-a+b\right)\left(r-a-b\right)\right]^{17/2}} \left[ 5 r^{22} + 5 r^{20} \left(a^2+b^2\right) \right. \nonumber \\
    & & -5 r^{18} \left(23 a^4-67 a^2 b^2+23 b^4\right)+5 r^{16} \left(a^2+b^2\right) \left(53 a^4-143 a^2 b^2+53 b^4\right) \nonumber \\
    & & -4 r^{14} \left(26 a^8+515 a^6 b^2-1350 a^4 b^4+515 a^2 b^6+26 b^8\right) \nonumber \\
    & & -4 r^{12} \left(a^2+b^2\right) \left(98 a^8-1365 a^6 b^2+2780 a^4 b^4-1365 a^2 b^6+98 b^8\right) \nonumber \\
    & & +2 r^{10} \left(308 a^{12}-1386 a^{10} b^2-4482 a^8 b^4+12165 a^6 b^6-4482 a^4 b^8-1386 a^2 b^{10}+308 b^{12}\right) \nonumber \\
    & & -10 r^8 \left(a^2+b^2\right) \left(32 a^{12}+150 a^{10} b^2-1698 a^8 b^4+3081 a^6 b^6-1698 a^4 b^8+150 a^2 b^{10}+32 b^{12}\right)\nonumber \\
    & & -5 r^6 \left(a^2-b^2\right)^2 \left(a^{12}-438 a^{10} b^2+3 a^8 b^4+2436 a^6 b^6+3 a^4 b^8-438 a^2 b^{10}+b^{12}\right) \nonumber \\
    & & +r^4 \left(a^2-b^2\right)^4 \left(a^2+b^2\right) \left(59 a^8-228 a^6 b^2-2602 a^4 b^4-228 a^2 b^6+59 b^8\right) \nonumber \\
    & & -r^2 \left(a^2-b^2\right)^6 \left(13 a^8+181 a^6 b^2+396 a^4 b^4+181 a^2 b^6+13 b^8\right) \nonumber \\
    & & \left. -\left(a^2-b^2\right)^8 \left(a^2+b^2\right) \left(a^4+5 a^2 b^2+b^4\right) \right]~.
\end{eqnarray}
\end{widetext}
As expected, the disk-disk attraction in Eq.~(\ref{eq:int_DD_attr_final}) and repulsion in Eq.~(\ref{eq:int_DD_rep_final}) are reduced to the disk-point attraction in Eq.~(\ref{eq:int_DP_attr_final}) and repulsion in Eq.~(\ref{eq:int_DP_rep_final}), respectively, in the limit of $b \rightarrow 0$ under the constraint of $\pi b^2 \lambda_b = 1$. Furthermore, both attraction and repulsion between two disks are symmetric between $a$ and $b$.

\subsection{Wall Potentials for LJ Point Particles and Disks in 2D}

To confine a LJ liquid and a disk in 2D, it is useful to derive the relevant wall potentials. The LJ potential can be integrated for a point mass interacting with a half-plane filled with uniformly distributed LJ material points. The resulting point-wall potential in 2D is
\begin{equation}
    \label{eq:int_PW_final}
    U_\text{PW}(r) = \frac{3\pi\lambda_w}{32}\left( \frac{21A_2}{80r^{10}} - \frac{A_1}{r^4} \right)~,
\end{equation}
where $\lambda_w$ is the areal density of LJ point masses that make up the half-plane and $r$ is the distance of the point mass from the boundary of the half-plane (i.e., the wall).

The disk-point potential can be integrated for a disk and a half-plane to produce the disk-wall potential,
\begin{widetext}
\begin{eqnarray}
    \label{eq:int_DW_final}
    U_\text{DW}^\text{R}(r) = \frac{3\pi^2\lambda_w \lambda_a a^2 r}{32} \left[ \frac{21 A_2 (35 a^6+280a^4r^2 +336a^2r^4+64r^6)}{5120(r^2-a^2)^{17/2}} -\frac{A_1}{(r^2-a^2)^{5/2}} \right]~.
\end{eqnarray}
\end{widetext}
Here $r$ is the distance from the disk's center to the wall. In the limit of $a \rightarrow 0$ under the constraint of $\pi a^2\lambda_a = 1$, the disk-wall potential is reduced to the point-wall potential in Eq.~(\ref{eq:int_PW_final}). This is consistent with the fact that the disk-wall potential can also be obtained by integrating the point-wall potential for a disk and a wall.

\section{Simulation Methods}\label{sec:disk:md_methods}

The integrated potentials presented in the previous section can be used for a wide range of 2D systems containing disks. They are implemented in the LAMMPS MD simulator\cite{LAMMPS} as a user package termed \texttt{DISK} (see the Supplementary Material for details). Then LAMMPS is used to perform all the simulations reported here on 2D suspensions of disks in an explicit solvent. The equation of motion is integrated with a velocity Verlet algorithm using a timestep of $\delta t = 0.01\tau$. Various thermalization schemes based on a Nose-Hoover thermostat,\cite{Nose1984, Hoover1984} a Langevin thermostat, or a dissipative particle dynamics (DPD) thermostat\cite{Groot1997JCP} are utilized to control temperature. Comparisons and discussions of thermostats are included in the Supplementary Material. For simulations under a constant pressure, a Nos\'{e}-Hoover barostat is used.\cite{Martyna1994}

\begin{figure}[htb]
    \centering
    \includegraphics[width=0.45\textwidth]{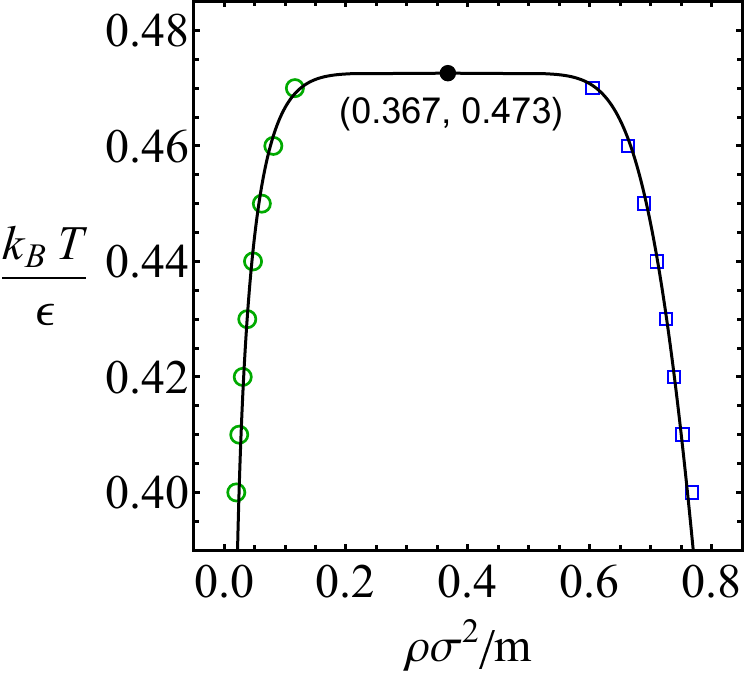}
    \caption{Coexisting densities vs. temperature for the LJ liquid in 2D. The lines are the fits based on Eqs.~(\ref{eq:den_fit_one}) and (\ref{eq:den_fit_two}). The black dot indicates the critical point determined from such fits.}
    \label{fig:phase_diagram}
\end{figure}

When simulating colloidal suspensions, the solvent is frequently modeled as a LJ liquid consisting of point particles that interact with each other through the potential in Eq.~(\ref{eq:lj_potential}).\cite{Cheng2013} Compared with their three-dimensional counterparts, 2D liquids exhibit stronger thermal fluctuations due to a reduced number of neighboring particles, which lead to lower critical temperatures and more pronounced interfacial effects.\cite{Homes2025} To understand the phase behavior of LJ liquids in 2D, $10^6$ point particles are placed in a rectangular simulation box with dimensions of $800 \sigma \times 3200\sigma$. The LJ interaction between solvent particles is truncated at $3\sigma$. A weak Langevin thermostat with a damping constant of $\Gamma = 0.01 \tau^{-1}$ is applied to all particles in the system with a target temperature ranging from $0.35 \epsilon/k_\text{B}$ to $0.5\epsilon/k_\text{B}$. After the liquid-vapor equilibrium is established at each temperature, a liquid slab is formed with two distinct liquid-vapor interfaces, which in 2D appear as fluctuating lines along the $x$-axis. A visualization of such system is included the Supplementary Material. The coexisting liquid ($\rho_L$) and vapor ($\rho_V$) densities are calculated at each temperature. The results are shown in Fig.~\ref{fig:phase_diagram}.

The liquid and vapor coexisting densities are found to follow the scaling form,\cite{Cheng2011JCP}
\begin{equation}
    \label{eq:den_fit_one}
    \rho_L + \rho_V = a - b T~,
\end{equation}
and
\begin{equation}
    \label{eq:den_fit_two}
    \rho_L - \rho_V = A \left( 1 - T/T_c \right)^{\beta}~,
\end{equation}
where $a$, $b$, and $A$ are fitting parameters and $T_c$ is the critical temperature. The critical exponent is fixed to the value of the 2D Ising universality class, $\beta = 0.125$.\cite{Onsager1944, Yang1952} For the 2D LJ liquid studied here with a cutoff of $3\sigma$, the critical point is $T_c \simeq 0.473\epsilon/k_\text{B}$ and $\rho_c \simeq 0.367 m/\sigma^2$.

% Figure on disk wetting
\begin{figure}[htb]
    \centering
    \includegraphics[width=0.48\textwidth]{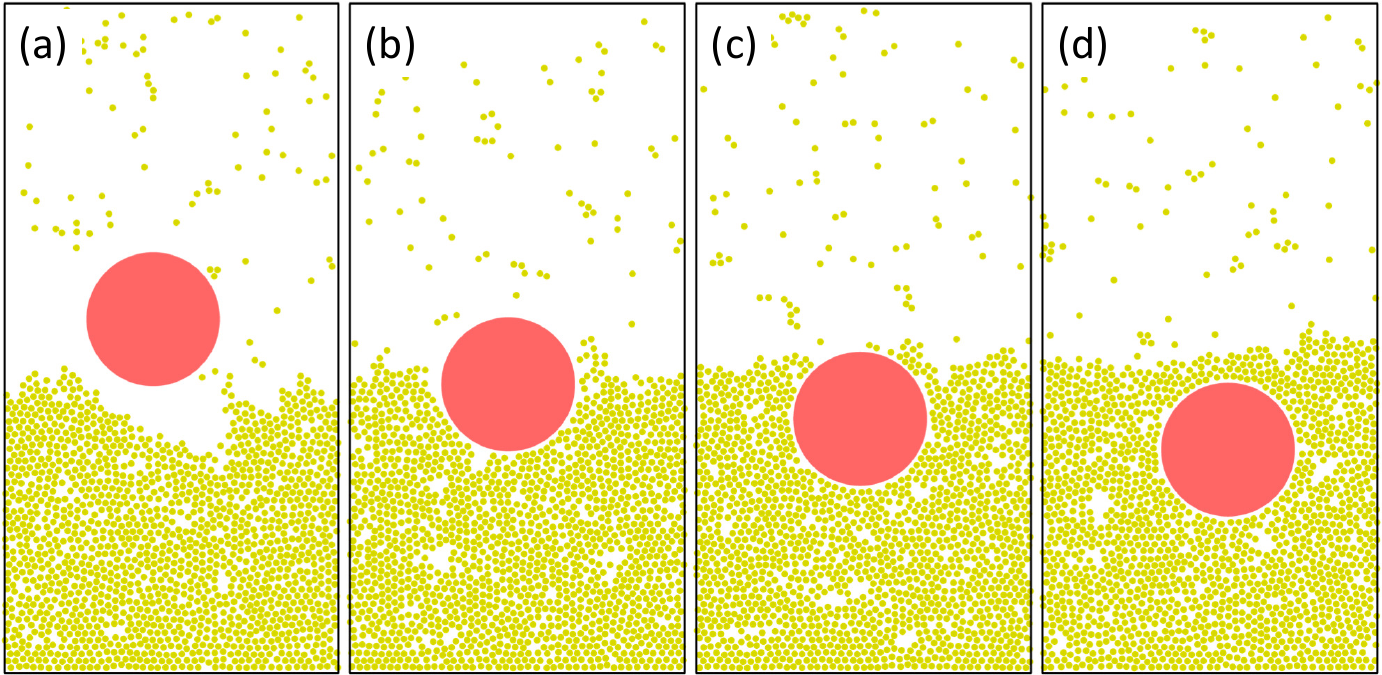}
    \caption{Wettability of a single disk of $10\sigma$ radius by the 2D LJ liquid at $T=0.42\epsilon/k_\text{B}$ and different Hamaker constants (a) $A_\text{ds} = 2\epsilon$ ($\epsilon_\text{ds} = 0.5\epsilon$), (b) $A_\text{ds} = 4\epsilon$ ($\epsilon_\text{ds} = 1.0\epsilon$), (c) $A_\text{ds} = 6\epsilon$ ($\epsilon_\text{ds} = 1.5\epsilon$), and (d) $A_\text{ds} = 8\epsilon$ ($\epsilon_\text{ds} = 2.0\epsilon$).}
    \label{fg:2D_disk_wetting}
\end{figure}

The 2D LJ liquid at temperature $T=0.42\epsilon/k_\text{B}$ is used as a solvent to prepare disk suspensions. To ensure that the disks are fully solvated, the wettability of a disk with $10\sigma$ radius by the LJ liquid in 2D is investigated. The interaction between the disk and a solvent particle is given by Eqs.~(\ref{eq:int_DP_attr_final}) and (\ref{eq:int_DP_rep_final}) with $A_1 = A_\text{ds}\sigma^6$ and $A_2 = A_\text{ds}\sigma^{12}$, respectively, where $A_\text{ds}$ is a Hamaker constant setting the interaction strength. The visualizations at various values of $A_\text{ds}$ are shown in Fig.~\ref{fg:2D_disk_wetting}. Note that $A_\text{ds} = 4 \epsilon_\text{ds}$, where $\epsilon_\text{ds}$ is the strength of the LJ potential being integrated to yield the disk-point potential governing the disk-solvent interaction. The results in Fig.~\ref{fg:2D_disk_wetting} show that when $\epsilon_\text{ds}$ is increased to about $2\epsilon$, the disk is fully wetted by the solvent. In the following simulations, $\epsilon_\text{ds}$ is set at $2\epsilon$ to ensure that the disks are well dispersed in the LJ solvent to form a uniform dispersion at equilibrium.

Below we report MD simulations of disk suspensions, where either monodisperse or bidisperse disks are dispersed in an explicit solvent at various area fractions. A disk interacts with nearby solvent particles via the integrated disk-point potential with a strength set by a Hamaker constant $A_\text{ds} = 8\epsilon$ and a cutoff of $R+4\sigma$, where $R$ is the disk radius. To ensure the well-dispersion of disks, the disk-disk potential is truncated at its minimum to make the direct disk-disk interaction purely repulsive. For example, for disks with $R=10\sigma$, the disk-point potential is truncated at $14\sigma$ while the cutoff of the disk-disk potential is $r_c = 10.6083\sigma$. The Hamaker constant setting the strength of the disk-disk potential is $A_\text{dd} = 4\epsilon$.

\section{Results and Discussion}\label{sec:disk:res}

\subsection{Equilibrium Structure of Monodisperse Disk Suspensions}\label{sec:disk:mono_suspension}

\begin{figure*}[htb]
    \centering
    \includegraphics[width=\textwidth]{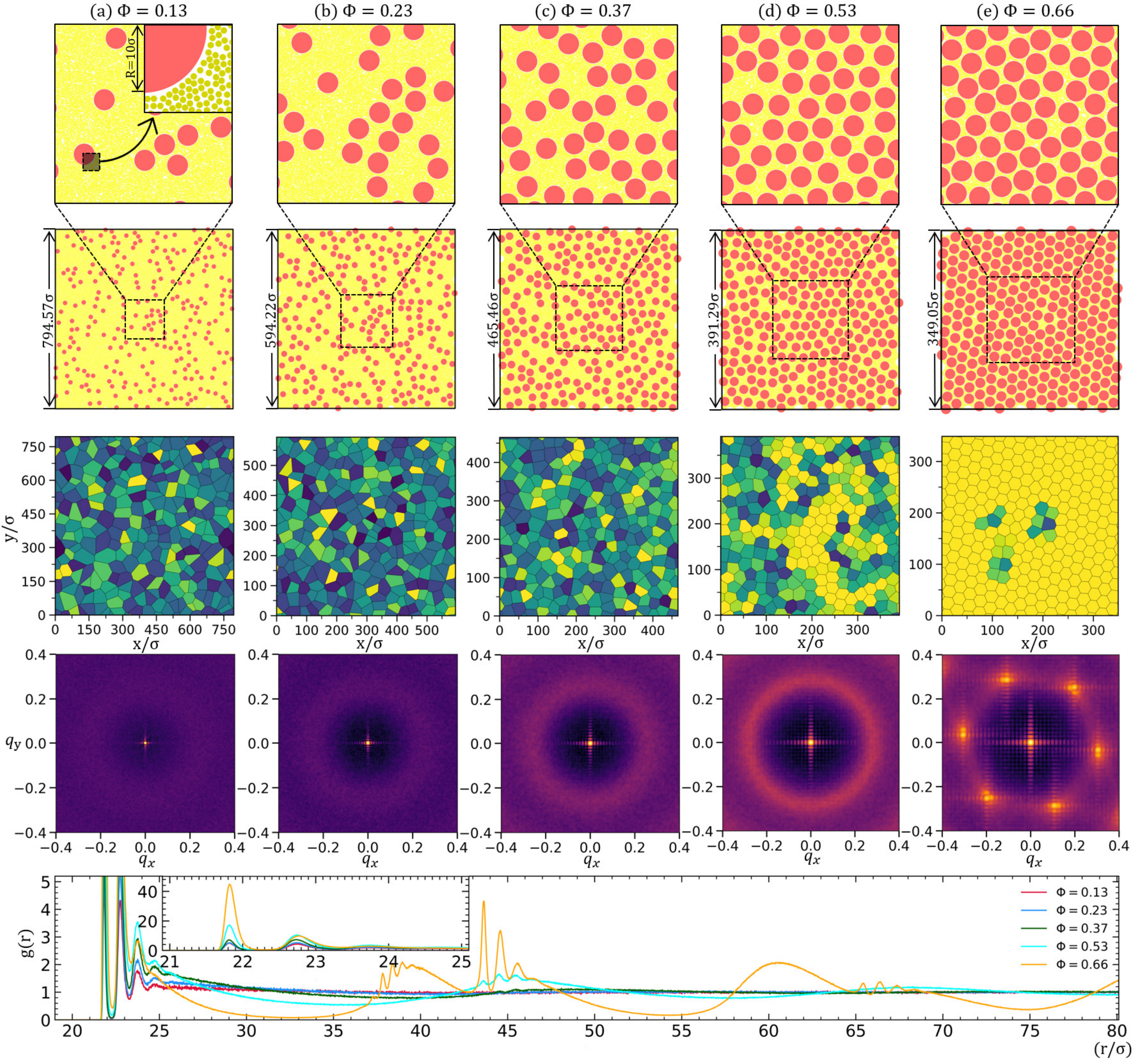}
    \caption{\textbf{First and second rows}: Visualizations of disk suspensions at increasing area fractions ($\Phi$). The disk radius is $10\sigma$ and the number of disks is fixed at $256$. The number of solvent particles (i.e., LJ point masses) is tuned to realize different area fractions. Full systems are visualized in the second row, with a square of dimensions of $175\sigma \times 175\sigma$ at the center of each system magnified in the first row. The inset of the leftmost image of the first row illustrates the size ratio of the disk and a LJ particle. \textbf{Third row}: Voronoi tesselation analysis of the corresponding configurations of disks in the second row. Yellow cells are used to highlight disks with a high value of the orientational order parameter, i.e., $|\Phi_{6}(i)| \gtrsim 0.8$, which is a characteristic of the hexagonal packing. \textbf{Fourth row}: Static structure factor $S(q_x, q_y)$ in the reciprocal space, revealing the emergence and sharpening of Bragg peaks as $\Phi$ is increased. The rightmost panel clearly shows the six-fold symmetry of the hexagonal packing of disks at $\Phi=0.66$. \textbf{Fifth row}: Disk-disk radial distribution function, $g(r)$, for all suspensions for $r$ up to $80\sigma$. The inset shows the features of $g(r)$ at short ranges.}
    \label{fig:disk_suspensions_equi}
\end{figure*}

The integrated potentials and the simulation methods established in the previous sections are used to investigate the equilibrium structure of disk suspensions in 2D. In the first system, 256 disks of a radius of $R=10\sigma$ are inserted into a LJ solvent uniformly filling a square box with dimensions of $800\sigma \times 800\sigma$. The disks are randomly placed in such a way that their overlaps are avoided. Any solvent particles that overlap with a placed disk are removed. Then the system is equilibrated at a target temperature of $T=0.42\epsilon/k_\text{B}$ and a target pressure of $P = 0.05 \epsilon/\sigma^3$ controlled with a Nos\'{e}-Hoover thermostat\cite{Nose1984, Hoover1984} and barostat.\cite{Martyna1994} The equilibration run lasts $5\times10^{4}\tau$. The simulation box for the equilibrated system remains a square but the side length is reduced to $794.6\sigma$. The corresponding area fraction of disks is $\Phi \simeq 0.13$. The equilibrated system is then used for production runs under a constant area. The temperature of the solvent is set at $T=0.42\epsilon/k_\text{B}$ with a weak Langevin thermostat with a damping rate of $0.01\tau^{-1}$, which is coupled to all the solvent particles in the system. It is confirmed that the results presented below for equilibrium disk suspensions are insensitive to the type of thermostat used.

After the equilibration run, half of the solvent particles are removed to prepare the second suspension with a larger area fraction. The new system is further equilibrated under a constant pressure and temperature and the procedure of removing solvent particles is repeated to prepare successive systems with the number of disks fixed at 256, resulting in an increasing area fraction of the disks. In total, five systems are prepared with the following area fractions: $0.13$, $0.23$, $0.37$, $0.53$, and $0.66$. The corresponding simulation box sizes are $794.6\sigma$, $594.2\sigma$, $465.5\sigma$, $391.3\sigma$, and $349.1\sigma$, respectively.

Once each system is equilibrated, a production run of $2\times 10^5\tau$ is conducted under a constant area. Visualizations of the equilibrium suspensions are included in Fig.~\ref{fig:disk_suspensions_equi}. In the second row of Fig.~\ref{fig:disk_suspensions_equi}, the entire system of each suspension is visualized. Note that the size of the simulation box decreases as the area fraction of disks is increased. In the first row of Fig.~\ref{fig:disk_suspensions_equi}, a square region of a fixed side length of $175\sigma$ at the center of each simulation box is visualized, with an inset illustrating the contrasting sizes of the disks and the LJ solvent particles. All these visualizations show that at low area fractions, the disks are dispersed randomly. However, as the area fraction is increased, disks start to pack with an emerging local order. At $\Phi \simeq 0.66$, the hexagonal packing of disks is clearly visible.

To quantitatively characterize the disk distribution in each suspension, Voronoi tessellation analysis is performed for the configurations of disks shown in the second row of Fig.~\ref{fig:disk_suspensions_equi}. A Voronoi diagram provides insight into the local packing and coordination environment of each disk. For example, if a disk is surrounded by six nearest neighbors, then the corresponding Voronoi cell is a hexagon. The results from the Voronoi partition are included in the third row of Fig.~\ref{fig:disk_suspensions_equi}. To quantify the packing order, a local orientational order parameter ($\Phi_6$) is computed using the Voronoi diagrams. Specifically, for the $i$-th disk, $\Phi_{6}(i)$ is defined as
\begin{equation}
    \Phi_6(i) = \frac{1}{N_b(i)} \sum_{j=1}^{N_{b}(i)} e^{i6\theta_{ij}}~,
    \label{eq:orientation_order_para}
\end{equation}
where $N_b(i)$ is the number of neighboring disks of the $i$-th disk, $j$ runs through all the neighboring disks, and $\theta_{ij}$ is the angle between a predefined axis and the position vector pointing from the center of the $i$-th disk to the center of its $j$-th neighbor. $\Phi_6$ therefore characterizes the degree of six-fold symmetry in the local neighborhood surrounding a given disk. Its magnitude, $|\Phi_{6}|$, lies between 0 and 1, with values close to 1 indicating a high degree of hexagonal packing, while lower values signaling a more disordered distribution of disks. Each Voronoi cell in Fig.~\ref{fig:disk_suspensions_equi} is colored with a scale set by $|\Phi_6|$ of the corresponding disk that the cell belongs to. For cells with $|\Phi_{6}| \gtrsim 0.8$, a bright yellow color is used while darker colors are assigned to cells with lower values of $|\Phi_{6}|$. The results unequivocally show the emergence of hexagonal packing structure, signaling a disorder-order transition, as the area fraction of disks is increased.

The fourth row of Fig.~\ref{fig:disk_suspensions_equi} shows the 2D static structure factor, $S(\mathbf{q}) \equiv S(q_x, q_y)$, of the corresponding distribution of disks shown in the second row of Fig.~\ref{fig:disk_suspensions_equi}. The structure factor provides a statistical description of the packing of disks in reciprocal space and is analogous to the diffraction pattern observed in a scattering experiment. $S(\mathbf{q})$ is calculated as follows,
\begin{equation}
    S(\mathbf{q}) = \frac{1}{N} \left| \sum_{j=1}^{N} e^{-i \, \mathbf{q} \cdot \mathbf{r}_j} \right|^2~,
    \label{eq:structure_factor}
\end{equation}
where $N$ is the total number of disks, $\mathbf{q} \equiv (q_x, q_y)$ is the two-dimensional scattering vector, and $\mathbf{r}_j$ is the position vector of the center of the $j$-th disk in real space.

The 2D density plots of the structure factor, $S(q_x, q_y)$, shown in Fig.~\ref{fig:disk_suspensions_equi} are obtained by averaging over 100 statistically independent configurations. At low area fractions, $S(q_x, q_y)$ exhibits a diffuse ring around a characteristic value of $|q|$, which corresponds to the average distance between a disk and its nearest neighbor. Otherwise, the structure factor is rather featureless. As the area fraction of disks is increased, the ring sharpens. At $\Phi=0.53$, the ring becomes prominent, reflecting the coexistence of crystal-like domains, where the disks exhibit hexagonal packing, and liquid-like domains, where the distribution of disks is more disordered. At this area fraction, the characteristic separation between nearest neighboring disks is well defined. In the system with $\Phi=0.66$, the disks are almost all hexagonally packed everywhere. As a result, the corresponding plot of $S(q_x, q_y)$ exhibit sharp Bragg peaks on the ring set by the characteristic value of $|q|$. The arrangement of these peaks displays a distinct six-fold symmetry, which is the definitive signature of a long-range hexagonal crystal structure in 2D.

Results on the radial distribution function of disks, $g(r)$, are included in the last row of Fig.~\ref{fig:disk_suspensions_equi}. These functions provide another way to quantify the structural features of the distribution of disks revealed by the Voronoi tessellation analysis and the static structure factor. At all area fractions, $g(r)$ exhibits a series of peaks between $r=20\sigma$ and $25\sigma$ (see the inset in the last row of Fig.~\ref{fig:disk_suspensions_equi}), corresponding to one, two, and three layers of solvent particles in the gap between two adjacent disks, which have a radius of $R=10\sigma$. This is because the direct disk-disk interaction is purely repulsive but there is an effective attraction between disks mediated by the solvent. As the area fraction is increased, additional peaks of $g(r)$ emerge at disk-disk separations around $4R$, $6R$, and $8R$, indicating that the disks start to pack into an ordered structure. Such spatially separated set of peaks are especially visible at $\Phi=0.66$, where the disks form a hexagonal lattice.

\begin{figure}[htb]
    \centering
    \includegraphics[width=0.45\textwidth]{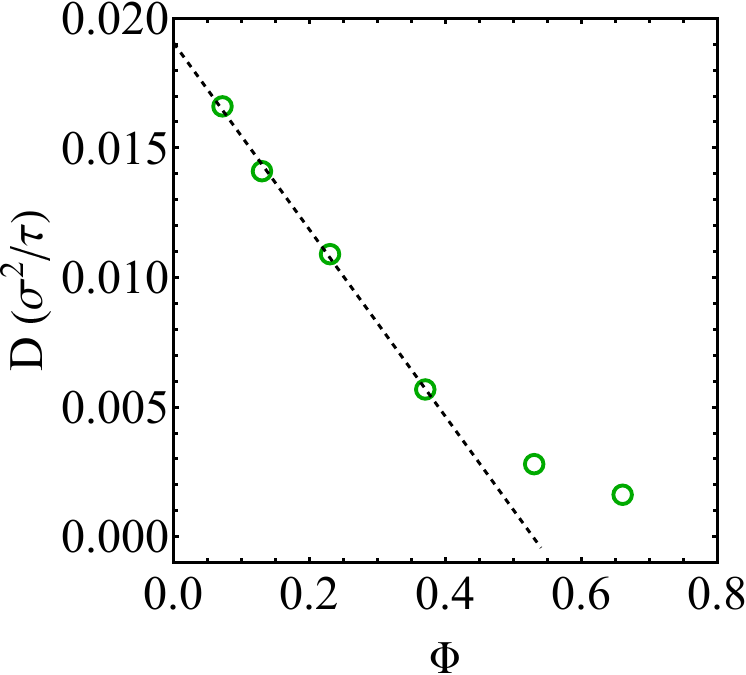}
    \caption{Diffusion coefficient of disks, $D$, determined from the mean-square displacement is plotted against their area fraction, $\Phi$. A linear fit (dashed line) to the data from four systems where the disks are disorderly packed yields $D = D_0+\chi\Phi$ with $D_0= 1.91\times 10^{-2}\sigma^2/\tau$ and $\chi = -3.61\times 10^{-2}\sigma^2/\tau$.}
    \label{fig:diff_coeff_area}
\end{figure}

The mean-square displacement, $\langle \Delta \mathbf{r}^2(t)\rangle$, of disks is calculated as a function of time interval $t$ and averaged over individual disks and different starting times with independent simulations of disk suspensions at various area fractions. Then, $\lim_{t\rightarrow \infty} \frac{\langle \Delta \mathbf{r}^2(t)\rangle}{4t}$ is taken as the effective diffusion coefficient ($D$) of the disks, though it is known that in 2D, the velocity autocorrelation function has a long time tail and $\langle \Delta \mathbf{r}^2(t)\rangle \sim t \sqrt{\ln t}$.\cite{Choi2017NJP} In the simulations performed here for calculating $D$, a weak Langevin thermostat with a damping rate of $0.01\tau^{-1}$ is applied to all the solvent particles in each system, As a result, hydrodynamic interactions are suppressed and $\langle \Delta \mathbf{r}^2(t)\rangle$ scales linearly with $t$. That is, $\lim_{t\rightarrow \infty} \frac{\langle \Delta \mathbf{r}^2(t)\rangle}{4t}$ reaches a plateau for a sufficiently large $t$, which allows $D$ to be reasonably estimated. The results on $D$ are plotted against $\Phi$ in Fig.~\ref{fig:diff_coeff_area}. As expected, $D$ decreases with an increasing $\Phi$. The data from four systems with small $\Phi$, where the disk packing is disordered, can be fit to $D = D_0+\chi \Phi$ with $D_0= 1.91\times 10^{-2}\sigma^2/\tau$ and $\chi = -3.61\times 10^{-2}\sigma^2/\tau$. At $\Phi = 0.53$ and $0.66$, hexagonal crystalline domains of disks emerge and the diffusion coefficient deviates from the linear fit. This is expected as the linear fit would indicate that $D$ is reduced to 0 at $\Phi \simeq 0.53$. However, the maximum area fraction of (monodisperse) disk packing in 2D is about 0.91. The suspensions studied here are well below this limit. Consequently, the disks remain in a fluid-like state with significant freedom for Brownian motion and long-range diffusion.

%section on evaporading monodisperse disk suspensions
\subsection{Evaporation of Monodisperse Disk Suspensions}

\begin{figure*}[htb]
    \centering
    \includegraphics[width=\textwidth]{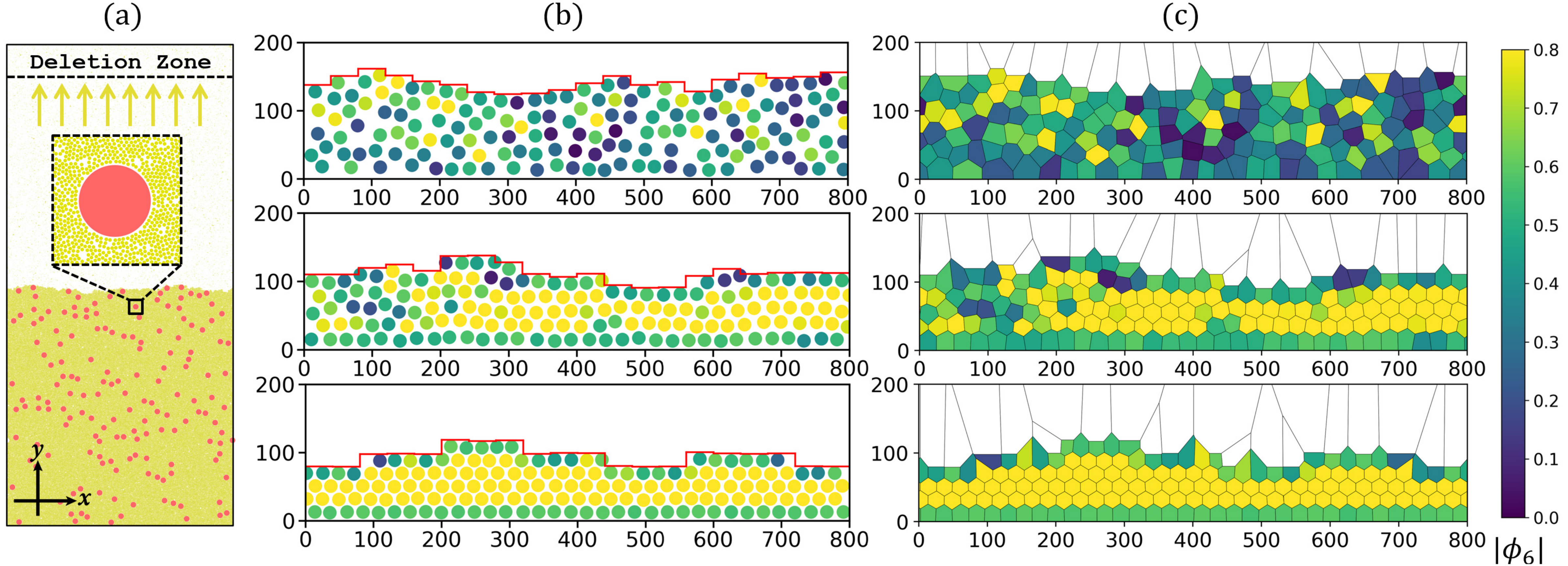}
    \caption{(a) Visualization of the initial state of a 2D monodisperse suspension of disks prior to evaporation. The width of the suspension is $800\sigma$ and its initial thickness is about $834\sigma$. The solvent particles are colored yellow while the disks are colored red. (b) Evolution of the disk distribution in the drying suspension at a receding speed of $v_e \simeq 1.8\times 10^{-4} \sigma/\tau$ for the liquid-vapor interface. The elapsed time since the initiation of evaporation for each frame from top to bottom is $4 \times 10^{6}\tau$, $4.2 \times 10^{6} \tau$, and $4.4 \times10^{6} \tau$, respectively. For clarity, the solvent is not shown. Each disk is assigned a color based on its value of $|\Phi_{6}|$ computed with neighboring disks determined via Voronoi tessellation analysis. Each red stepped line indicates the surface of the drying film of disks. (c) Voronoi tessellations of the disk distributions in (b).}
    \label{fig:mono_disk_evap}
\end{figure*}

To further test the disk potentials, we prepare a 2D monodisperse suspension of disks of $R=10\sigma$. The entire system contains 482,949 solvent particles and 162 disks. All the interaction parameters are identical to those used in Sec.~\ref{sec:disk:mono_suspension}. Periodic boundary conditions are applied along the $x$-axis. In the $y$ direction, the system is confined from below with a wall at $y=0$ that is attractive to the solvent but repulsive to the disks. A top wall at $y=1700\sigma$, which is repulsive to both solvent and disks, is used to confine the system from above. The wall potentials for a solvent particle and a disk are given in Eqs.~(\ref{eq:int_PW_final}) and (\ref{eq:int_DW_final}), respectively, with $A_1 = 4\epsilon\sigma^6$, $A_2 = 4\epsilon\sigma^{12}$, and $\lambda_w = \lambda_a = 1.0\sigma^{-2}$. The system is well equilibrated at $T = 0.42 \epsilon/k_\text{B}$ prior to evaporation, with the disks uniformly dispersed in the liquid solvent that is at equilibrium with its vapor (see Fig.~\ref{fig:mono_disk_evap}(a)). The initial thickness of the suspension is $834\sigma$ and that of the vapor phase is $866\sigma$ based on the average location of the liquid-vapor interface. The initial area fraction of disks is about $0.076$. To induce evaporation, a deletion zone with thickness $100\sigma$ is introduced at the top of the vapor zone, where solvent vapor particles are randomly removed at a fixed rate. For the simulation of the drying process reported here, one solvent particle is removed every $10\tau$, which makes the liquid-vapor interface to recede at a speed of $v_e \simeq 1.8\times 10^{-4} \sigma/\tau$. Therefore, the P\'{e}clet number of the evaporating system, defined as $\text{Pe} \equiv \frac{H v_e}{D}$ with $H$ being the initial thickness of the suspension, is about $9$. In this simulation, the temperature of the solvent liquid and vapor is maintained at $T=0.42\epsilon/k_\text{B}$ using a DPD thermostat with a friction coefficient of $\gamma=1.0m/\tau$ and a cutoff of $r_c = 3.0\sigma$ (see the Supplementary Material regarding the choice of thermostat).

As the solvent evaporates, the liquid-vapor interface recedes and the disks become more concentrated in the liquid suspension. As shown in Fig.~\ref{fig:mono_disk_evap}(b), the distribution of disks remains disordered in the early stage of drying. When the thickness of the suspension film is reduced to about $100\sigma$, the area fraction of disks has increased to about $0.6$ and domains of hexagonally packed disks emerge and coexist with disordered domains. It is noted that the film of packed disks exhibits a rough surface at the evaporation front. In the final film shown in Fig.~\ref{fig:mono_disk_evap}(b), all the disks except those adjacent to the bottom wall or at the surface are in hexagonal packing. The film still exhibits a piecewise stepped surface.

To characterize the structural evolution in the suspension during drying, we perform Voronoi tessellation analysis of the disk distribution. Such analysis starts with a set of points at the centers of the disks. Periodic images of the disks along the $x$-axis are used to determine the Voronoi cells for the disks near the lateral boundary of the simulation box. However, special care is needed for the disks near the bottom wall and the receding liquid-vapor interface. To this end, the simulation box is divided into 20 long bins along the $x$-axis with a width of $40\sigma$ for each bin. The topmost disk in each bin with the largest $y$ coordinate is identified. Its coordinates are denoted as $(x_t, y_t)$. Then a ghost disk with its center located at $(x_t, y_t+2R)$ is added to the list of disks used for Voronoi tessellation. Furthermore, the intersection between the horizontal line located at $y=y_t+R$ and the spatial bin being considered is taken as the top surface of the packed film in this bin. At the bottom wall which is a sharp, flat boundary, all the disks in the suspension are mirrored with respect to the $x$-axis at $y=0$. The mirror images are used to determine the lower boundary of the Voronoi cells of the disks near the bottom wall.

The Voronoi tessellations of the disk distributions in Fig.~\ref{fig:mono_disk_evap}(b) are shown in Fig.~\ref{fig:mono_disk_evap}(c). From such analysis, the neighbors of each disk are identified and the information is then used to compute the local orientational order parameter, $\Phi_6$, for the disk under consideration through Eq.~(\ref{eq:orientation_order_para}). A scale based on $|\Phi_6|$ is then used to color the Voronoi cells in Fig.~\ref{fig:mono_disk_evap}(c) and the disks in Fig.~\ref{fig:mono_disk_evap}(b), with yellow indicating a local hexagonal arrangement. The results clearly show the emergence and development of hexagonal disk packing during the drying process.

The analysis described above also yields the top surface of the drying film, which is shown as a stepped red line in each panel of Fig.~\ref{fig:mono_disk_evap}(b). The rough edge of the film persists throughout the drying process, which reflects the significant fluctuation effects inherent in 2D fluids. As the number of nearest neighbors of a particle is reduced, the 2D suspension is more prone to fluctuations and the liquid-vapor interfaces recedes in a more irregular manner. As a result, the packed film of disks in 2D induced by drying develops a stepped edge at the evaporation front.

\subsection{Evaporation of Bidisperse Disk Suspensions}

\begin{figure}[htb]
    \centering
    \includegraphics[width=0.45\textwidth]{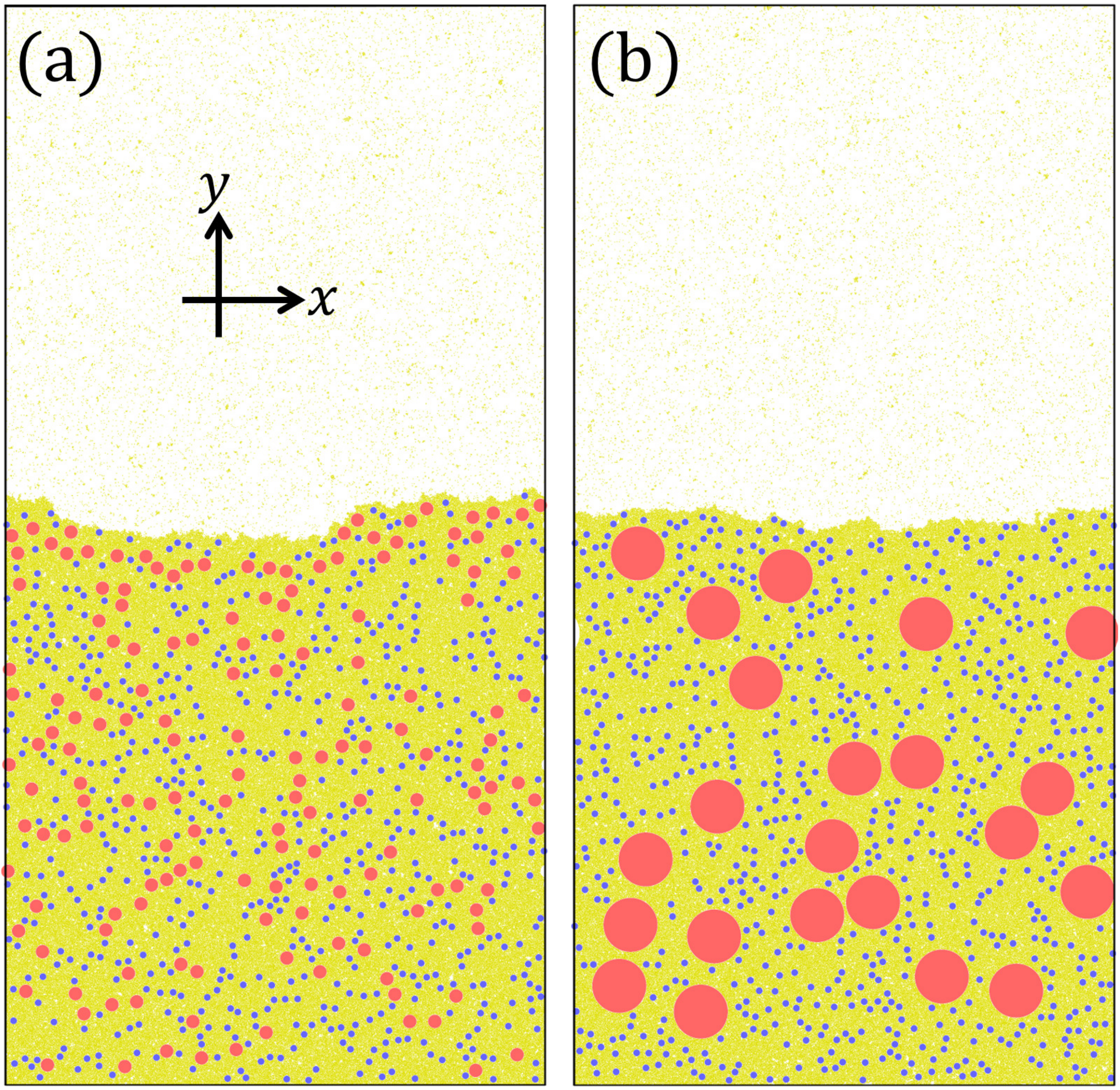}
    \caption{Bidisperse disk suspensions with $R_L/R_S=2$ (left) and 8 (right). In each system, the large disks are colored red while the small disks are colored blue.}
    \label{fig:binary_susp_sys}
\end{figure}

\begin{figure*}[htb]
    \centering
    \includegraphics[width=0.9\textwidth]{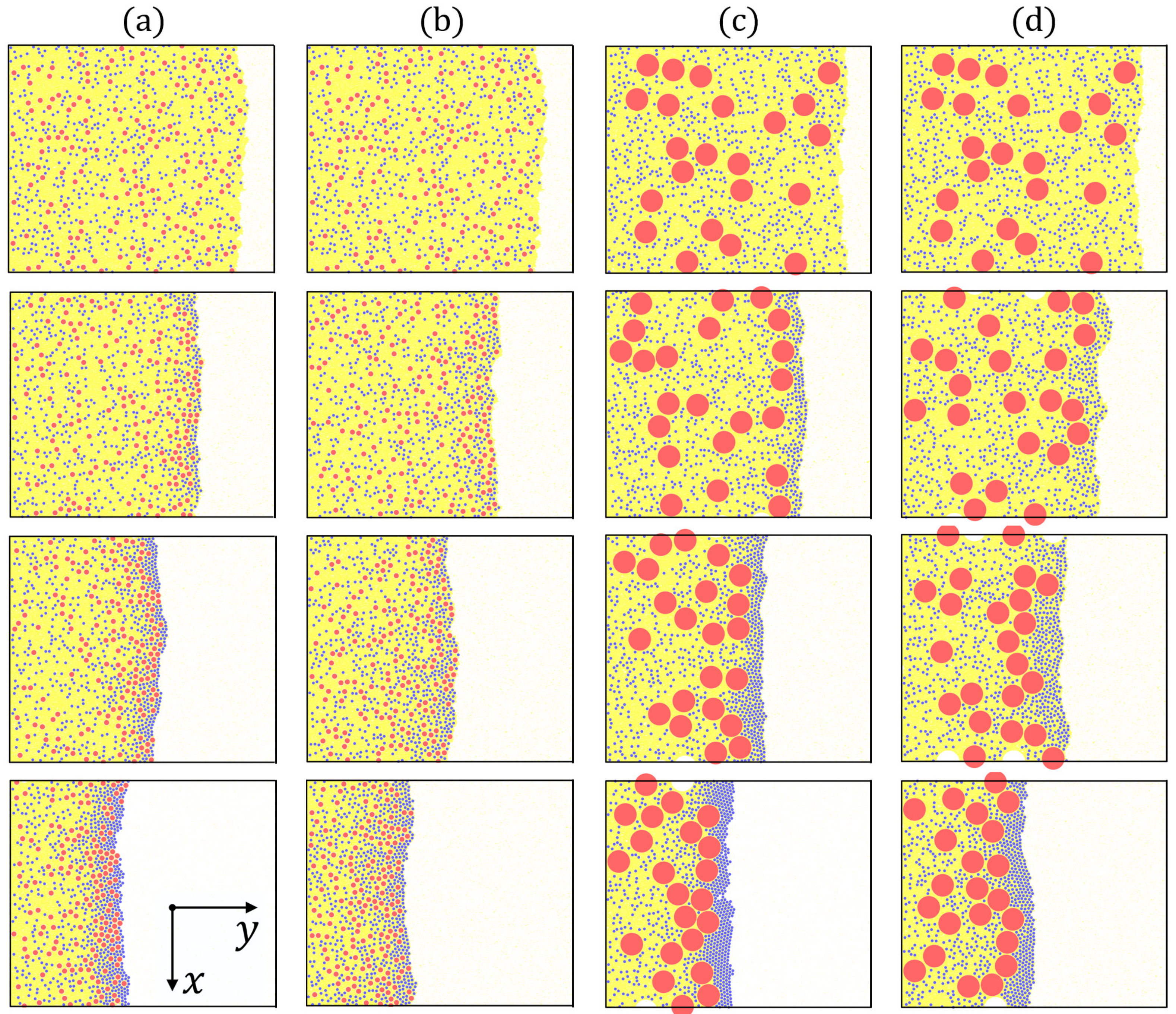}
    \caption{Visualizations of bidisperse disk suspensions with $R_L/R_S=2$ [(a) and (b)] and $R_L/R_S=8$ [(c) and (d)] under fast [(a) and (c)] and slow [(b) and (d)] drying. In each set, the four visualizations from top to bottom are at the following elapsed times since the initiation of drying: (a) $0$, $0.45\times 10^{6}\tau$, $0.9\times 10^{6}\tau$, and $1.35\times 10^{6}\tau$ ; (b) $0$, $0.9\times 10^{6}\tau$, $1.8\times 10^{6}\tau$, and $2.7\times 10^{6}\tau$; (c) $0$, $0.42\times 10^{6}\tau$, $0.84\times 10^{6}\tau$, and $1.26\times 10^{6}\tau$; (d) $0$, $0.77\times 10^{6}\tau$, $1.54\times 10^{6}\tau$, and $2.31\times 10^{6}\tau$.}
    \label{fig:binary_susp_drying_visual}
\end{figure*}

Polydisperse colloidal suspensions have recently attracted wide attention as when they are rapidly dried, the colloidal particles developed stratified distributions.\cite{Fortini2016, Zhou2017} In the past decade, the stratification phenomena are extensively studied in three-dimensional systems. An interesting question is whether they occur in polydisperse 2D suspensions as well. The disk potentials presented here provide an opportunity to address this question. To this end, the suspension model used in the previous section is adapted to create two bidisperse suspensions, each containing a mixture of large and small disks dispersed in the 2D solvent at $T=0.42 \epsilon/k_\text{B}$. The radius of the small disk ($R_S$) is fixed at $5\sigma$. The radius of the large disk ($R_L$) is $10\sigma$ in one suspension but is increased to $40\sigma$ in the other. Therefore, the size ratio $R_L/R_S$, is 2 and 8, respectively. All the interactions, including those with the confining walls, are implemented similarly as in the monodisperse suspensions discussed above. The details on the interactions among the solvent particles and the disks can also be found in Sec.~\ref{sec:disk:md_methods}. The bidisperse suspension with $R_L/R_S=2$ consists of $433,751$ solvent atoms, $162$ large disks, and $641$ small disks. The initial thickness of the suspension is about $839\sigma$. The initial area fraction is $0.076$ for the large disks and $0.075$ for the small disks. The suspension with $R_L/R_S=8$ consists of $381,436$ solvent atoms, $20$ large disks, and $641$ small disks. The initial thickness of this suspension is about $840\sigma$. The initial area fraction is $0.15$ for the large disks and $0.075$ for the small disks. The two suspensions after long equilibration runs are visualized in Fig.~\ref{fig:binary_susp_sys}. As expected, the two types of disks are all uniformly dispersed in the liquid solvent. Furthermore, the liquid-vapor interface, which sets the edge of the suspension phase, exhibits clear roughness in both systems.

\begin{figure*}[htb] % Notice the asterisk (*) here
    \centering
    \includegraphics[width=\textwidth]{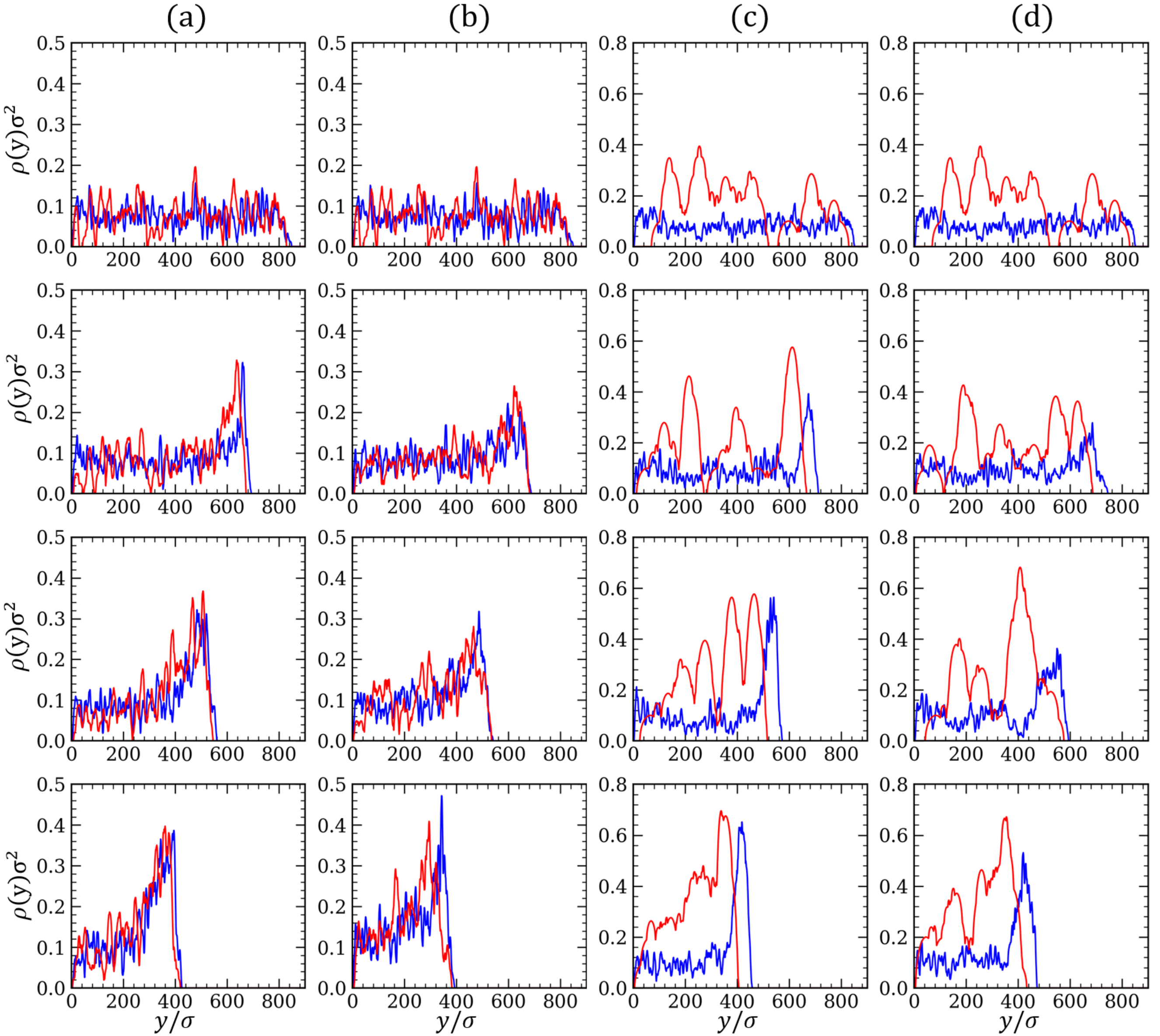}
    \caption{Evolution of the local area fraction, ${\rho}(y)$, of disks in bidisperse disk suspensions with $R_L/R_S=2$ [(a) and (b)] and $R_L/R_S=8$ [(c) and (d)] under fast [(a) and (c)] and slow [(b) and (d)] drying. In each plot, results for large disks are shown in red and those for small disks are shown in blue. In each set, the four panels from top to bottom are at the following elapsed times since the initiation of drying: (a) $0$, $0.45\times 10^{6}\tau$, $0.9\times 10^{6}\tau$, and $1.35\times 10^{6}\tau$ ; (b) $0$, $0.9\times 10^{5}\tau$, $1.8\times 10^{6}\tau$, and $2.7\times 10^{6}\tau$; (c) $0$, $0.42\times 10^{6}\tau$, $0.84\times 10^{6}\tau$, and $1.26\times 10^{6}\tau$; (d) $0$, $0.77\times 10^{6}\tau$, $1.54\times 10^{6}\tau$, and $2.31\times 10^{6}\tau$. That is, each panel here has a one-to-one correspondence with the panel in Fig.~\ref{fig:binary_susp_drying_visual}.}
    \label{fig:binary_susp_drying_den}
\end{figure*}

Solvent evaporation is implemented in an identical way as in the evaporating monodisperse suspension described in the previous section. A deletion zone of a thickness of $100\sigma$ is introduced to the top of the vapor zone and right below the top confining wall. Solvent vapor particles in this zone are removed at a preset rate. Two drying rates are simulated for each suspension. Under fast drying, one solvent vapor particle is removed every $5\tau$, making the liquid-vapor interface to recede at a speed of $v_e \simeq 4.0\times 10^{-4}\sigma/\tau$ in the suspension with $R_L/R_S=2$. A slow rate, which is half of the fast rate, is realized by doubling the time interval of removing one solvent vapor particle to $10\tau$. Under the slow rate, the liquid-vapor interface recedes at a speed of $v_e \simeq 2.0\times 10^{-4}\sigma/\tau$. For the suspensions with $R_L/R_S=8$, the receding speed of the interface is slightly higher, with $v_e \simeq 2.2\times 10^{-4}\sigma/\tau$ for slow drying and $v_e \simeq 4.4\times 10^{-4}\sigma/\tau$ for fast drying. This is because at $R_L/R_S=8$, the total area fraction of disks has to be higher to ensure that a sufficient number of large disks can be included in the suspension. The effective diffusion coefficients ($D$) of disks are calculated from the mean-square displacement with independent simulations of bidisperse suspensions. For the suspension with $R_L/R_S=2$, such calculations yield $D_{L} \simeq 1.33\times 10^{-2}\sigma^{2}/\tau$ for the disks with $10\sigma$ radius and $D_{S} \simeq 1.94\times 10^{-2}\sigma^{2}/\tau$ for the disks with $5\sigma$ radius. In this suspension, the corresponding P\'{e}clet numbers are $\text{Pe}_{L}\simeq 25 $ and $\text{Pe}_{S}\simeq 17$ under fast drying and $\text{Pe}_{L}\simeq 13 $ and $\text{Pe}_{S}\simeq 9$ under slow drying. For the suspension with $R_L/R_S=8$, the large disks have a larger radius of $40\sigma$ and diffuse more slowly with $D_{L} \simeq 4.2\times 10^{-3}\sigma^{2}/\tau$. The diffusion coefficient of the small disks ($R_{s}=5\sigma$) is $D_{S} \simeq 1.75\times 10^{-2}\sigma^{2}/\tau$, which slightly lower than that of the same-sized disks in the system with $R_L/R_S=2$. This is because the latter system has an overall lower area fraction of disks. For the suspension with $R_L/R_S=8$, the P\'{e}clet numbers are $\text{Pe}_{L}\simeq 88$ and $\text{Pe}_{S}\simeq 21$ under fast drying and $\text{Pe}_{L}\simeq 44$ and $\text{Pe}_{S}\simeq 11$ under slow drying.

The two bidisperse disk suspensions in the various stages of drying under either slow or fast solvent evaporation rates are visualized in Fig.~\ref{fig:binary_susp_drying_visual}. For each suspension, the results at the two drying rates are qualitatively similar. In the suspension with $R_L/R_S=2$, large disks are found to accumulate at the evaporation front first in the early stage of drying, especially under the slow-drying condition. Then small disks quickly catch up and accumulate near the receding interface as well. As more solvent is evaporated, small disks outnumber large disks at the evaporation front. In the final drying suspension visualized, there is a slightly relative enrichment of small disks over large disks at the top of the drying film, indicating ``small-on-top'' stratification that is similar to the outcome in three-dimensional polydisperse colloidal suspensions being quickly dried.\cite{Fortini2016, Zhou2017, Martin-Fabiani2016, Makepeace2017, Tang2018Langmuir, Schulz2018, Schulz2021, Palmer2023, Carr2018, LiuXiao2018, LiuWeiping2019, Carr2024, Dong2020SM, Sofroniou2024, Cusola2018, Romermann2019, Coureur2023, Hooiveld2023, Hooiveld2025, LiSiyu2023, Tiwari2023}

Stratification is more significant in the suspension with $R_L/R_S=8$. Under both solvent evaporation rates, small disks begin to concentrate immediately in a region below the receding liquid-vapor interface as the drying process starts. This enrichment zone widens with the progress of drying. Below this zone enriched with small disks, large disks are accumulated. The distribution of disks in the drying film clearly shows a stratified structure. The results thus show that stratification is enhanced when the size ratio of disks in the mixture is increased. A similar trend is also observed in three-dimensional suspensions.\cite{Fortini2016}

The distribution of disks in a drying film is quantified by computing the local area fraction of disks along the direction of drying. To this end, a simulation box is divided into consecutive bins of $2.0\sigma$ width along the $y$-axis. The area of a disk is distributed to all the bins it intersects with based on the area overlap with each bin. Then the local area fraction of disks in a bin is calculated and assigned to the center of the bin. That is, a local area fraction, $\rho(y)$, represents the fraction of a bin spanning $y - 1.0 \sigma$ and $y + 1.0 \sigma$ along the $y$-axis that is occupied by disks. In each suspension, the local area fraction is calculated at different times during drying for both small and large disks.

Results on the evolution of $\rho(y)$ for large and small disks are included in Fig.~\ref{fig:binary_susp_drying_den}. They corroborate the analysis presented above. The suspension with $R_L/R_S=2$ only exhibits a weak sign of stratification upon drying. The degree of stratification is slightly enhanced under the slow-drying condition. where the solvent evaporation rate is half the rate in fast drying. However, strong stratification is observed in the suspension with $R_L/R_S=8$ at both drying rates. Below the receding interface, there is a layer of small disks where large disks are depleted. The area fraction of small disks shows a significant peak in this layer and the peak value increases with time as the film thins during drying. Furthermore, at similar extent of drying, the peak value is higher at the larger evaporation rate of the solvent. Below the surface layer of small disks, concentrated large disks coexist with small disks. The area fraction profile of large disks shows a negative gradient towards the bulk of the suspension. The area fraction of large disks has similar peak values under both fast and slow drying conditions. The results thus indicate that for a disk mixture with a larger size disparity, stratification is slightly stronger when the solvent evaporates faster.

The stratification phenomena in the 2D drying suspensions of bidisperse disks reported in this paper indicate that a similar physical mechanism is in operation as in three-dimensional suspensions.\cite{Fortini2016, Zhou2017, Sear2017, Sear2018, Howard2017, Statt2017, Tang2018Langmuir, Schulz2021, Rees-Zimmerman2021, Rees-Zimmerman2024, Liu2025JCP} During rapid drying, the evaporation-induced advection of disks outplay their diffusion and as a result, disks are accumulated at the evaporation front and develop concentration gradients that gradually decrease into the bulk suspension. These concentration gradients drive the suspended disks into diffusiophoretic motion. However, in a bidisperse disk mixture the effect is asymmetric. The diffusiophoretic mobility of the larger disks driven by the concentration gradient of the smaller ones is larger than that of the smaller disks induced by the concentration gradient of the larger disks. The net result is that the larger disks are driven out of the region near the evaporation front, which is populated and enriched with the smaller disks. The outcome of the combination of fast drying and diffusiophoresis is the so-called ``small-on-top'' stratification, as seen here in the bidisperse suspensions of disks that are dried quickly. As the solvent evaporation rate is further reduced, ``large-on-top'' stratified or uniform distributions are expected to emerge, which remain to be confirmed with more studies in the future.

\section{Conclusions}\label{sec:disk:conc}

In this study, we derive analytical potentials for disks in two dimensions, including the disk-point, disk-disk, and disk-wall potentials. By treating the disk and the wall as a uniform distribution of Lennard-Jones material points, we integrate the Lennard-Jones potential to obtain the analytical forms of the disk potentials. All the results are validated by comparing with the results from numerical integrations to ensure their accuracy. These potentials are implemented in the LAMMPS simulator and used to study two-dimensional suspensions of disks with an explicit solvent modeled as a Lennard-Jones liquid.

In monodisperse disk suspensions, the distribution of disks transitions from a disordered state to a hexagonal packing as the area fraction of disks is increased. A similar transition is also observed when the solvent evaporates out of a monodisperse disk suspension. As the drying process proceeds, the disk concentration in the suspension increases and a hexagonal arrangement emerges in the drying film. However, the drying film has a persistent rough edge where the disk packing exhibit visible steps.

Finally, suspensions of bidisperse disk mixtures with two size ratios are studied. The ``small-on-top'' stratification phenomenon, where the small disks form a concentrated layer at the evaporation front and the large disks are enriched below this surface layer, is observed after the suspensions are rapidly dried. The simulation results show that the degree of stratification is much stronger for larger size ratios. However, the effect of the solvent evaporation rate on stratification seems to depend on the size ratio of the disks in the mixture and more studies are still needed to fully understand the role of drying rate in 2D colloidal suspensions.

%\section*{Supplementary Material}
%See the Supplementary Material for details on the following:
%\begin{itemize}
%\item Detailed derivations of the analytical forms of the integrated potentials involving disks in two dimensions.
%\item Details on the implementation of the analytical potentials of disks in LAMMPS as a user package (\texttt{DISK}).
%\item Comparison of Langevin and DPD thermostats for simulating drying of disk suspensions in two dimensions.
%\end{itemize}

\section*{Acknowledgements}
This material is based on work supported by the National Science Foundation under Grant No. DMR-1944887. This research was initially supported by a 4-VA Collaborative Research Grant (``Material Fabrication via Droplet Drying''). This work was performed, in part, at the Center for Integrated Nanotechnologies, an Office of Science User Facility operated for the U.S. Department of Energy Office of Science. Sandia National Laboratories is a multimission laboratory managed and operated by National Technology and Engineering Solutions of Sandia, LLC., a wholly owned subsidiary of Honeywell International, Inc., for the U.S. Department of Energy's National Nuclear Security Administration under contract DE-NA0003525. This paper describes objective technical results and analysis. Any subjective views or opinions that might be expressed in the paper do not necessarily represent the views of the U.S. Department of Energy or the United States Government.

% references
%\bibliography{disk-disk}

%%%%%%%%%%%%%%%%%%%%%%%%%%%%%%%%%%%%%%%%%%%%%%%%%%%%%%%%%
% Supporting Information
\clearpage
\newpage
\onecolumngrid
\renewcommand{\thesection}{S\arabic{section}}
\setcounter{section}{0}
\renewcommand{\thefigure}{S\arabic{figure}}
\setcounter{figure}{0}
\renewcommand{\theequation}{S\arabic{equation}}
\setcounter{equation}{0}
\renewcommand{\thepage}{SI-\arabic{page}}
\setcounter{page}{1}
\begin{center}
{\bf \large Supplementary Information for ``Analytical Interaction Potentials for Disks in Two Dimensions''}
\end{center}
\begin{center}
{\bf Binghan Liu$^{1,2,3}$, Junwen Wang$^{2,3,4}$, Gary S. Grest$^{5}$, and Shengfeng Cheng$^{1,2,3,4}$}
\end{center}
\begin{center}
{$^1$Department of Physics, $^2$Center for Soft Matter and Biological Physics, $^3$Macromolecules Innovation Institute, and $^4$Department of Mechanical Engineering, Virginia Tech, Blacksburg, Virginia 24061, USA\\$^5$Sandia National Laboratories, Albuquerque, NM 87185, USA}
\end{center}

\section{Theoretical Model of Integrated Lennard-Jones Potentials}

In this section, we include detailed derivations of the integrated Lennard-Jones potentials involving disks in two dimensions (2D), including integrated disk-point, disk-disk, disk-wall, and point-wall potentials.

The interaction between two point particles is governed by the Lennard-Jones (LJ) 12-6 potential, 
\begin{equation}
    \label{eq:lj_potential}
    U_\text{LJ}(r) = 4\epsilon \left[\left( \frac{\sigma}{r}\right)^{12} - \left(\frac{\sigma}{r}\right)^6 \right]~,
\end{equation}
where $\epsilon$ and $\sigma$ are the energy and length scales of the interaction, and $r$ is the inter-particle distance. For simplicity, the LJ potential can be rewritten as
\begin{equation}
    \label{eq:lj_potential_two}
    U_\text{LJ}(r) = \frac{A_2}{r^{12}} - \frac{A_1}{r^{6}}~,
\end{equation}
where $A_2 = 4\epsilon \sigma^{12}$ and $A_1 = 4\epsilon \sigma^{6}$. Therefore, $\epsilon$ and $\sigma$ can be obtained from $A_1$ and $A_2$ via $\epsilon = A_1^2/(4 A_2)$ and $\sigma = \left( A_2 /A_1 \right)^{1/6}$.

\subsection{Integrated Disk-Point Potential}

\begin{figure}[htb]
    \centering
    \includegraphics[width=0.5\textwidth]{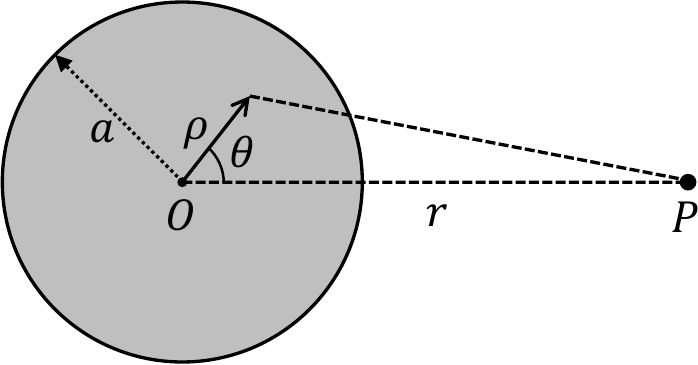}
    \caption{A polar coordinate system describing a 2D configuration of a disk of radius $a$ and a mass point $P$.}
    \label{fg:disk-point-geo}
\end{figure}

Using the coordinate system defined in Fig.~\ref{fg:disk-point-geo}, the integrated interaction between a disk of radius $a$ and a mass point at a distance of $r$ from the disk's center can be written as
\begin{eqnarray}
    \label{eq:int_DP}
    U_\text{DP}(r) &=& \lambda_a \int_{0}^{a} \int_{0}^{2\pi}  U_\text{LJ}(x)\vert_{x=\sqrt{r^2 + \rho^2 -2 r \rho \cos\theta}}~ \rho \text{d}\rho \text{d}\theta~,
\end{eqnarray}
where $\lambda_a$ is the areal density of LJ material points makign up the disk. Here, $\lambda_a$ is assumed to be constant, and $r>a$ is also expected.

The integrated disk-point potential can be separated to two terms, attraction and repulsion. The attractive terms reads
\begin{eqnarray}
    \label{eq:int_DP_attr}
    U_\text{DP}^\text{A}(r) &=& -A_1 \lambda_a \int_{0}^{a} \int_{0}^{2\pi}  \frac{\rho}{\left(r^2 + \rho^2 -2 r \rho \cos\theta\right)^3} \text{d}\rho \text{d}\theta~,
\end{eqnarray}
and the repulsive term reads
\begin{eqnarray}
    \label{eq:int_DP_rep}
    U_\text{DP}^\text{R}(r) &=& A_2 \lambda_a \int_{0}^{a} \int_{0}^{2\pi}  \frac{\rho}{\left(r^2 + \rho^2 -2 r \rho \cos\theta\right)^6} \text{d}\rho \text{d}\theta~,
\end{eqnarray}
It is convenient to define the following integral,
\begin{equation}
    \label{eq:Phi_func_def}
    \Phi_k(t) = \int_{0}^{2\pi} \frac{1}{\left(t-\cos\theta\right)^k} \text{d}\theta~.
\end{equation}
The expressions for $\Phi_k(t)$ with $k=3,4,...,10$ are
\begin{equation}
    \label{eq:Phi_func_three}
    \Phi_3(t) = \frac{\pi \left(2 t^2+1\right)}{\left(t^2-1\right)^{5/2}}~.
\end{equation}
\begin{equation}
    \label{eq:Phi_func_four}
    \Phi_4(t) = \frac{\pi t \left(2 t^2+3\right)}{\left(t^2-1\right)^{7/2}}~.
\end{equation}
\begin{equation}
    \label{eq:Phi_func_five}
    \Phi_5(t) = \frac{\pi \left( 8 t^4+24 t^2+3 \right) }{4 \left(t^2-1\right)^{9/2}}~.
\end{equation}
\begin{equation}
    \label{eq:Phi_func_six}
    \Phi_6(t) = \frac{\pi t \left(8 t^4+40 t^2+15\right)}{4 \left(t^2-1\right)^{11/2}}~.
\end{equation}
\begin{equation}
    \label{eq:Phi_func_seven}
    \Phi_7(t) = \frac{\pi \left(16 t^6+120 t^4+90 t^2+5\right)}{8 \left(t^2-1\right)^{13/2}}~.
\end{equation}
\begin{equation}
    \label{eq:Phi_func_eight}
    \Phi_8(t) = \frac{\pi t \left(16 t^6+168 t^4+210 t^2+35\right)}{8 \left(t^2-1\right)^{15/2}}~.
\end{equation}
\begin{equation}
    \label{eq:Phi_func_nine}
    \Phi_9(t) = \frac{\pi \left(128 t^8+1792 t^6+3360 t^4+1120 t^2+35\right)}{64 \left(t^2-1\right)^{17/2}}~.
\end{equation}
\begin{equation}
    \label{eq:Phi_func_ten}
    \Phi_{10}(t) = \frac{\pi t \left(128 t^8+2304 t^6+6048 t^4+3360 t^2+315\right)}{64 \left(t^2-1\right)^{19/2}}~.
\end{equation}

The integral in Eq.~(\ref{eq:int_DP_attr}) can be rewritten as
\begin{eqnarray}
    \label{eq:int_DP_attr_two}
    U_\text{DP}^\text{A}(r) &=& -A_1 \lambda_a \int_{0}^{a} \frac{\rho d\rho}{\left(2 r \rho\right)^3} \int_{0}^{2\pi}  \frac{1}{\left(t- \cos\theta\right)^3} \text{d}\theta~ \nonumber \\
    &=& -A_1 \lambda_a \int_{0}^{a} \frac{\rho}{\left(2 r \rho\right)^3} \Phi_3(t) \text{d}\rho \nonumber \\
    &=& -\pi A_1 \lambda_a \int_{0}^{a} \frac{\rho}{\left(2 r \rho\right)^3} \frac{\left(2 t^2+1\right)}{\left(t^2-1\right)^{5/2}}  \text{d}\rho
\end{eqnarray}
where $t=\frac{r^2 + \rho^2}{2 r \rho}$. Therefore,
\begin{eqnarray}
    \label{eq:int_DP_attr_three}
    U_\text{DP}^\text{A}(r) &=& -2\pi A_1 \lambda_a \int_{0}^{a}  \frac{\rho\left(r^4 + 4 r^2\rho^2 +\rho^4 \right)}{\left(r^2-\rho^2\right)^{5}}  \text{d}\rho \nonumber \\
    &=& -\frac{\pi}{2} A_1 \lambda_a \frac{a^2\left(2 r^2 + a^2 \right)}{\left(r^2-a^2\right)^4}
\end{eqnarray}
The corresponding attractive force is
\begin{eqnarray}
    F_\text{DP}^\text{A}(r) &=& -\frac{\partial U_\text{DP}^\text{A} }{\partial r} \nonumber \\
    &=& 2 \pi A_1 \lambda_a a^2 \frac{r}{\left(r^2 -a^2 \right)^4} - 4 \pi A_1 \lambda_a a^2 \frac{\left(2r^2 +a^2 \right) r }{\left(r^2 -a^2 \right)^5} \nonumber \\
    &=& - 6 \pi A_1 \lambda_a a^2 \frac{\left(r^2 +a^2 \right) r }{\left(r^2 -a^2 \right)^5}
\end{eqnarray}
%Obviously, this force is always radial and attractive (i.e., negative), as expected.

Similarly, the integral in Eq.~(\ref{eq:int_DP_rep}) for the integrated disk-point repulsion can be rewritten as
\begin{eqnarray}
    \label{eq:int_DP_rep_two}
    U_\text{DP}^\text{R}(r) &=& A_2 \lambda_a \int_{0}^{a} \frac{\rho \text{d}\rho}{\left(2 r \rho \right)^6} \int_{0}^{2\pi}  \frac{1}{\left(t- \cos\theta\right)^6} \text{d}\theta~ \nonumber \\
    &=& A_2 \lambda_a \int_{0}^{a} \frac{\rho}{\left(2 r \rho\right)^6} \Phi_6(t) \text{d}\rho \nonumber \\
    &=& \frac{\pi}{4} A_2 \lambda_a \int_{0}^{a} \frac{\rho}{\left(2 r \rho\right)^6} \frac{t \left(8t^4 + 40t^2+15\right)}{\left(t^2-1\right)^{11/2}}  \text{d}\rho \nonumber \\
    &=& 2\pi A_2 \lambda_a \int_{0}^{a} \frac{\rho \left(r^{10}+25 r^8 \rho^2+100 r^6 \rho^4+100 r^4 \rho^6+25 r^2 \rho^8+\rho^{10} \right)}{\left(r^2-\rho^2\right)^{11}} \text{d}\rho \nonumber \\
    &=&  \frac{\pi}{5} A_2 \lambda_a \frac{a^2 \left(5 r^8 +40 a^2 r^6 +60 a^4 r^4 +20 a^6 r^2 + a^8\right)}{\left(r^2-a^2\right)^{10}}
\end{eqnarray}
The corresponding repulsive force is
\begin{eqnarray}
    F_\text{DP}^\text{R}(r) &=& -\frac{\partial U_\text{DP}^\text{R} }{\partial r} \nonumber\\
    &=&  - 8\pi A_2 \lambda_a a^2 \frac{\left(r^6 + 6 r^4 a^2 + 6 r^2 a^4 + a^6 \right) r}{\left( r^2 -a^2 \right)^{10}} + 4\pi A_2 \lambda_a a^2 \frac{\left(5 r^8 +40 a^2 r^6 +60 a^4 r^4 +20 a^6 r^2 + a^8\right) r }{\left( r^2 -a^2 \right)^{11}} \nonumber \\
    &=& 12\pi A_2 \lambda_a a^2 \frac{\left(r^8+10 r^6 a^2+20 r^4 a^4+10 r^2 a^6+a^8\right) r}{\left( r^2 -a^2 \right)^{11}}
\end{eqnarray}
%As expected, this force is always radial and repulsive (i.e., positive).

\subsection{Integrated Disk-Disk Potential}

\begin{figure}[htb]
    \centering
    \includegraphics[width=0.5\textwidth]{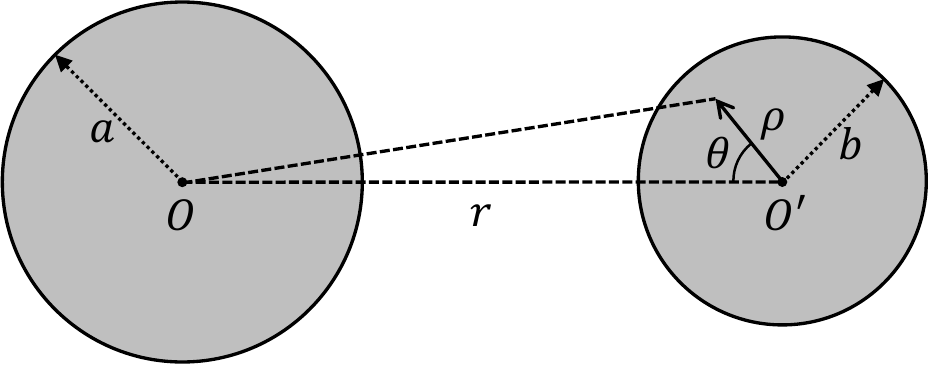}
    \caption{A polar coordinate system describing a 2D configuration of two disks of radius $a$ and $b$ at a center-to-center distance $r$.}
    \label{fg:disk-disk-geo}
\end{figure}

The interaction between two disks in two dimensions can be obtained by integrating the disk-point potential presented in the previous section over the second disk. Using the coordinate system defined in Fig.~\ref{fg:disk-disk-geo}, the integrated disk-disk potential is
\begin{eqnarray}
    \label{eq:int_DD}
    U_\text{DD}(r) &=& \lambda_b \int_{0}^{b} \int_{0}^{2\pi}  U_\text{DP}(x)\vert_{x=\sqrt{r^2 + \rho^2 -2 r \rho \cos\theta}}~ \rho \text{d}\rho \text{d}\theta~,
\end{eqnarray}
where $\lambda_b$, assumed to be constant, is the areal density of LJ material points on the second disk. Clearly, $r>a+b$ is assumed here. The integrated disk-point potential is
\begin{equation}
    U_\text{DP}(r) \equiv U_\text{DP}^\text{A}(r) + U_\text{DP}^\text{R}(r)
\end{equation}
with $U_\text{DP}^\text{A}(r)$ in Eq.~(\ref{eq:int_DP_attr_three}) and $U_\text{DP}^\text{R}(r)$ in Eq.~(\ref{eq:int_DP_rep_two}). Similarly, $U_\text{DD}(r) $ can be split into attractive and repulsive terms as
\begin{equation}
    U_\text{DD}(r) \equiv U_\text{DD}^\text{A}(r) + U_\text{DD}^\text{R}(r)
\end{equation}
with
\begin{eqnarray}
    \label{eq:int_DD_attr}
    U_\text{DD}^\text{A}(r) &=& \lambda_b \int_{0}^{b} \int_{0}^{2\pi}  U_\text{DP}^\text{A}(x)\vert_{x=\sqrt{r^2 + \rho^2 -2 r \rho \cos\theta}}~ \rho \text{d}\rho \text{d}\theta~,
\end{eqnarray}
and
\begin{eqnarray}
    \label{eq:int_DD_rep}
    U_\text{DD}^\text{R}(r) &=& \lambda_b \int_{0}^{b} \int_{0}^{2\pi}  U_\text{DP}^\text{R}(x)\vert_{x=\sqrt{r^2 + \rho^2 -2 r \rho \cos\theta}}~ \rho \text{d}\rho \text{d}\theta~,
\end{eqnarray}

\bigskip
\begin{center}
{\bf B.1. Integrated Disk-Disk Attraction}
\end{center}
\bigskip

To evaluate $U_\text{DD}^\text{A}(r)$, it is convenient to rewrite $U_\text{DP}^\text{A}(r)$ as 
\begin{eqnarray}
    \label{eq:int_DP_attr_split}
    U_\text{DP}^\text{A}(r) &=& -\frac{\pi a^2}{2} A_1 \lambda_a \left[ \frac{2}{\left(r^2-a^2\right)^3} + \frac{3a^2}{\left(r^2-a^2\right)^4}\right]
\end{eqnarray}
Then the attractive term in the integrated disk-disk potential is
\begin{eqnarray}
    \label{eq:int_DD_attr_two}
    U_\text{DD}^\text{A}(r) &=& -\frac{\pi a^2}{2}A_1\lambda_a\lambda_b \int_{0}^{b} \int_{0}^{2\pi} \left[ \frac{2}{\left(r^2+\rho^2-a^2-2r\rho \cos\theta \right)^3} + \frac{3a^2}{\left(r^2+\rho^2-a^2-2r\rho \cos\theta \right)^4}\right]\rho \text{d}\rho \text{d}\theta~.
\end{eqnarray}
With $t=\frac{r^2+\rho^2-a^2}{2r\rho}$, the first integration involved in Eq.~(\ref{eq:int_DD_attr_two}) can be written as
\begin{eqnarray}
    \label{eq:int_DD_attr_1st}
    & & \int_{0}^{b} \int_{0}^{2\pi} \frac{\rho}{\left(r^2+\rho^2-a^2-2r\rho \cos\theta \right)^3} \text{d}\rho \text{d}\theta \nonumber \\
    & =& \int_{0}^{b} \frac{\rho \text{d}\rho}{\left( 2r\rho\right)^3} \int_{0}^{2\pi} \frac{1}{\left(t-\cos\theta\right)^3}  \text{d}\theta \nonumber \\
    &=& \int_{0}^{b} \frac{\rho}{\left( 2r\rho\right)^3} \Phi_3(t)  \text{d}\rho \nonumber \\
      &=& \pi \int_{0}^{b} \frac{\rho}{\left( 2r\rho\right)^3} \frac{\left(2 t^2+1\right)}{\left(t^2-1\right)^{5/2}}  \text{d}\rho \nonumber \\
      &=& \pi \int_{0}^{b} \frac{\rho \left(2 \left( r^2+\rho^2-a^2\right)^2+(2r \rho)^2\right)}{\left(\left( r^2+\rho^2-a^2\right)^2-(2r \rho)^2\right)^{5/2}}  \text{d}\rho \nonumber \\
      &=& 2\pi \int_{0}^{b} \frac{\rho \left( \left( r^2+\rho^2-a^2\right)^2+2r^2 \rho^2\right)}{\left(\left( r^2+\rho^2-a^2\right)^2-4r^2 \rho^2\right)^{5/2}} \text{d}\rho
\end{eqnarray}
Similarly, the second integration in Eq.~(\ref{eq:int_DD_attr_two}) can be transformed as
\begin{eqnarray}
    \label{eq:int_DD_attr_2nd}
    & & \int_{0}^{b} \int_{0}^{2\pi} \frac{\rho}{\left(r^2+\rho^2-a^2-2r\rho \cos\theta \right)^4} \text{d}\rho \text{d}\theta \nonumber \\
    & =& \int_{0}^{b} \frac{\rho \text{d}\rho}{\left( 2r\rho\right)^4} \int_{0}^{2\pi} \frac{1}{\left(t-\cos\theta\right)^4}  \text{d}\theta \nonumber \\
    &=& \int_{0}^{b} \frac{\rho}{\left( 2r\rho\right)^4} \Phi_4(t)  \text{d}\rho \nonumber \\
      &=& \pi \int_{0}^{b} \frac{\rho}{\left( 2r\rho\right)^4} \frac{t\left(2 t^2+3\right)}{\left(t^2-1\right)^{7/2}}  \text{d}\rho \nonumber \\
      &=& \pi \int_{0}^{b} \frac{\rho \left( r^2+\rho^2-a^2\right) \left(2 \left( r^2+\rho^2-a^2\right)^2+3(2r \rho)^2\right)}{\left(\left( r^2+\rho^2-a^2\right)^2-(2r \rho)^2\right)^{7/2}}  \text{d}\rho \nonumber \\
      &=& 2\pi \int_{0}^{b} \frac{\rho \left( r^2+\rho^2-a^2\right) \left(\left( r^2+\rho^2-a^2\right)^2+6r^2 \rho^2\right)}{\left(\left( r^2+\rho^2-a^2\right)^2-4r^2 \rho^2 \right)^{7/2}}  \text{d}\rho       
\end{eqnarray}
Combining Eqs.~(\ref{eq:int_DD_attr_two}), (\ref{eq:int_DD_attr_1st}), and (\ref{eq:int_DD_attr_2nd}), we get
\begin{eqnarray}
    \label{eq:int_DD_attr_three}
    U_\text{DD}^\text{A}(r) &=& -\pi^2 A_1 a^2 \lambda_a\lambda_b \int_{0}^{b} \left[ \frac{2 \rho \left( \left( r^2+\rho^2-a^2\right)^2+2r^2\rho^2\right)}{\left(\left( r^2+\rho^2-a^2\right)^2-4r^2 \rho^2\right)^{5/2}} + \frac{3\rho a^2 \left( r^2+\rho^2-a^2\right) \left( \left( r^2+\rho^2-a^2\right)^2+6r^2 \rho^2\right)}{\left(\left( r^2+\rho^2-a^2\right)^2-4r^2 \rho^2\right)^{7/2}} \right] d\rho \nonumber \\
    &=& -\frac{\pi^2 a^2 b^2}{2} A_1 \lambda_a\lambda_b \frac{2r^4 - r^2 \left(a^2+b^2\right) - \left(a^2-b^2\right)^2}{\left[\left(r-a-b\right) \left(r-a+b\right) \left(r+a-b\right) \left(r+a+b\right)\right]^{5/2}}
\end{eqnarray} 
As expected, the expression of $U_\text{DD}^\text{A}(r)$ is symmetric when the parameters of the two disks are swapped. In the limit of $a\rightarrow 0$ and $b\rightarrow 0$ under the constraints of $\lambda_a \pi a^2=1$ and $\lambda_b \pi b^2=1$, each disk is reduced to a mass point and the integrated attraction is reduced to
\begin{equation}
    \lim_{a\rightarrow 0,~b\rightarrow 0} U_\text{DD}^\text{A}(r) = -\frac{A_1}{r^6}~,
\end{equation}
which is the LJ attraction between point particles, as expected. For two coplanar disks of the same radius $a$, the integrated attraction reads
\begin{equation}
    U_\text{DD}^\text{A}(r)|_{a=b} = -\pi^2 A_1 a^4 \lambda_a^2 \frac{r^2-a^2}{r^3\left( r^2-4a^2\right)^{5/2}}~.
\end{equation}

The attractive force between two disks can be computed as
\begin{eqnarray}
    F_\text{DD}^\text{A}(r) &=& - \frac{\partial U_\text{DD}^\text{A}}{\partial r} \nonumber \\
    &=& - 6 \pi^2 A_1 a^2 b^2 \lambda_a\lambda_b  \frac{\left[\left(r^2-a^2-b^2 \right)\left(r^2-a^2+b^2 \right)\left(r^2+a^2-b^2 \right) +2 r^2 a^2 b^2 \right]r}{
    \left[\left(r-a-b\right) \left(r-a+b\right) \left(r+a-b\right) \left(r+a+b\right)\right]^{7/2} }~.
\end{eqnarray}
When $a=b$, the expression of the attractive force becomes
\begin{equation}
    F_\text{DD}^\text{A}(r)|_{a=b} = - 6 \pi^2 A_1 a^4 \lambda_a^2  \frac{r^4 - 2 r^2 a^2 + 2a^4}{r^4 \left( r^2 - 4a^2\right)^{7/2}}~.
\end{equation}

\bigskip
\begin{center}
{\bf B.2. Integrated Disk-Disk Repulsion}
\end{center}
\bigskip

Similar strategies can be used to obtain the integrated repulsion, $U_\text{DD}^\text{R}(r)$, between two coplanar disks. To do this, we need to decompose $U_\text{DP}^\text{R}(r)$ as
\begin{eqnarray}
    \label{eq:int_DP_rep_split}
    U_\text{DP}^\text{R}(r) &=& \frac{\pi a^2}{5} A_2 \lambda_a \left[ \frac{5}{\left(r^2-a^2\right)^6} + \frac{60a^2}{\left(r^2-a^2\right)^7} + \frac{210a^4}{\left(r^2-a^2\right)^8} + \frac{280a^6}{\left(r^2-a^2\right)^9} + \frac{126a^8}{\left(r^2-a^2\right)^{10}} \right]
\end{eqnarray}
Each term in the square brackets in Eq.~(\ref{eq:int_DP_rep_split}) has to be integrated over the second disk with the help of $\Phi_k(t)$ in Eq.~(\ref{eq:Phi_func_def}).

Again, with $t=\frac{r^2+\rho^2-a^2}{2r\rho}$, the first integral can be evaluated as
\begin{eqnarray}
    \label{eq:int_DD_rep_1st}
    I_1 &=& \int_{0}^{b} \int_{0}^{2\pi} \frac{5\rho}{\left(r^2 +\rho^2 -a^2 - 2r\rho \cos\theta \right)^6} \text{d}\rho \text{d}\theta \nonumber \\
    &=& \int_{0}^{b} \frac{5\rho}{\left( 2r\rho\right)^6} \text{d}\rho \int_{0}^{2\pi} \frac{1}{\left(t - \cos\theta \right)^6} \text{d}\theta \nonumber \\
    &=& \int_{0}^{b} \frac{5\rho}{\left( 2r\rho\right)^6} \Phi_6(t) \text{d}\rho \nonumber \\
     &=& \int_{0}^{b} \frac{5\rho}{\left( 2r\rho\right)^6} \frac{\pi t \left( 8t^4 + 40t^2+15\right) }{4 \left( t^2-1\right)^{11/2}} \text{d}\rho~. 
\end{eqnarray}
With the expression of $t$ inserted, the first integral becomes
\begin{equation}
    \label{eq:int_DD_rep_1st_two}
    I_1 = \frac{5\pi}{4} \int_{0}^{b}  \frac{\rho h \left( 8 h^4 + 40h^2 \left( 2r\rho\right)^2+15 \left( 2r\rho\right)^4 \right) }{ \left( h^2-\left( 2r\rho\right)^2\right)^{11/2}} \text{d}\rho~,
\end{equation}
where $h \equiv r^2+\rho^2-a^2 = 2 r \rho t$.

The second integral reads
\begin{eqnarray}
    \label{eq:int_DD_rep_2nd}
    I_2 &= & \int_{0}^{b} \int_{0}^{2\pi} \frac{60a^2\rho}{\left(r^2 +\rho^2 -a^2 - 2r\rho \cos\theta \right)^7} \text{d}\rho \text{d}\theta \nonumber \\
    &=& \int_{0}^{b} \frac{60 a^2 \rho}{\left( 2r\rho\right)^7} \text{d}\rho \int_{0}^{2\pi} \frac{1}{\left(t - \cos\theta \right)^7} \text{d}\theta \nonumber \\
    &=& \int_{0}^{b} \frac{60 a^2 \rho}{\left( 2r\rho\right)^7} \Phi_7(t) \text{d}\rho \nonumber \\
     &=& \int_{0}^{b} \frac{60 a^2 \rho}{\left( 2r\rho\right)^7} \frac{\pi \left( 16 t^6 + 120t^4 + 90t^2+5\right) }{8 \left( t^2-1\right)^{13/2}} \text{d}\rho~.
\end{eqnarray}
With the expression of $t$ inserted, the second integral becomes
\begin{equation}
    \label{eq:int_DD_rep_2nd_two}
    I_2 = \frac{15\pi a^2}{2} \int_{0}^{b}  \frac{\rho \left( 16 h^6 + 120 h^4 \left( 2r\rho\right)^2 + 90h^2 \left( 2r\rho\right)^4 + 5 \left( 2r\rho\right)^6 \right) }{ \left( h^2-\left( 2r\rho\right)^2\right)^{13/2}} \text{d}\rho~.
\end{equation}

The third integral reads
\begin{eqnarray}
    \label{eq:int_DD_rep_3rd}
    I_3 &= & \int_{0}^{b} \int_{0}^{2\pi} \frac{210a^4\rho}{\left(r^2 +\rho^2 -a^2 - 2r\rho \cos\theta \right)^8} \text{d}\rho \text{d}\theta \nonumber \\
    &=& \int_{0}^{b} \frac{210 a^4 \rho}{\left( 2r\rho\right)^8} \text{d}\rho \int_{0}^{2\pi} \frac{1}{\left(t - \cos\theta \right)^8} \text{d}\theta \nonumber \\
    &=& \int_{0}^{b} \frac{210 a^4 \rho}{\left( 2r\rho\right)^8} \Phi_8(t) \text{d}\rho \nonumber \\
     &=& \int_{0}^{b} \frac{210 a^4 \rho}{\left( 2r\rho\right)^8} \frac{\pi t \left( 16 t^6 + 168t^4 + 210t^2+35\right) }{8 \left( t^2-1\right)^{15/2}} \text{d}\rho~.
\end{eqnarray}
With the expression of $t$ inserted, the third integral becomes
\begin{equation}
    \label{eq:int_DD_rep_3rd_two}
    I_3 = \frac{105\pi a^4}{4} \int_{0}^{b}  \frac{\rho h \left( 16 h^6 + 168 h^4 \left( 2r\rho\right)^2 + 210 h^2 \left( 2r\rho\right)^4 + 35 \left( 2r\rho\right)^6 \right) }{ \left( h^2-\left( 2r\rho\right)^2\right)^{15/2}} \text{d}\rho~.
\end{equation}

The fourth integral reads
\begin{eqnarray}
    \label{eq:int_DD_rep_4th}
    I_4&=& \int_{0}^{b} \int_{0}^{2\pi} \frac{280a^6\rho}{\left(r^2 +\rho^2 -a^2 - 2r\rho \cos\theta \right)^9} \text{d}\rho \text{d}\theta \nonumber \\
    &=& \int_{0}^{b} \frac{280 a^6 \rho}{\left( 2r\rho\right)^9} \text{d}\rho \int_{0}^{2\pi} \frac{1}{\left(t - \cos\theta \right)^9} \text{d}\theta \nonumber \\
    &=& \int_{0}^{b} \frac{280 a^6 \rho}{\left( 2r\rho\right)^9} \Phi_9(t) \text{d}\rho \nonumber \\
     &=& \int_{0}^{b} \frac{280 a^6 \rho}{\left( 2r\rho\right)^9} \frac{\pi \left( 128 t^8 + 1792 t^6 + 3360t^4 + 1120t^2+35\right) }{64 \left( t^2-1\right)^{17/2}} \text{d}\rho~.
\end{eqnarray}
With the expression of $t$ inserted, the fourth integral becomes
\begin{equation}
    \label{eq:int_DD_rep_4th_two}
    I_4 = \frac{35\pi a^6}{8} \int_{0}^{b}  \frac{\rho \left( 128 h^8 + 1792 h^6 \left( 2r\rho\right)^2 + 3360 h^4 \left( 2r\rho\right)^4 + 1120 h^2 \left( 2r\rho\right)^6 + 35 \left( 2r\rho\right)^8 \right) }{ \left( h^2-\left( 2r\rho\right)^2\right)^{17/2}} \text{d}\rho~.
\end{equation}

The fifth integral reads
\begin{eqnarray}
    \label{eq:int_DD_rep_5th}
    I_5&=& \int_{0}^{b} \int_{0}^{2\pi} \frac{126a^8\rho}{\left(r^2 +\rho^2 -a^2 - 2r\rho \cos\theta \right)^{10}} \text{d}\rho \text{d}\theta \nonumber \\
    &=& \int_{0}^{b} \frac{126 a^8 \rho}{\left( 2r\rho\right)^{10}} \text{d}\rho \int_{0}^{2\pi} \frac{1}{\left(t - \cos\theta \right)^{10}} \text{d}\theta \nonumber \\
    &=& \int_{0}^{b} \frac{126 a^8 \rho}{\left( 2r\rho\right)^{10}} \Phi_{10}(t) \text{d}\rho \nonumber \\
    &=& \int_{0}^{b} \frac{126 a^8 \rho}{\left( 2r\rho\right)^{10}} \frac{\pi t \left( 128 t^8 + 2304 t^6 + 6048t^4 + 3360t^2+315\right) }{64 \left( t^2-1\right)^{19/2}} \text{d}\rho~.
\end{eqnarray}
With the expression of $t$ inserted, the fifth integral becomes
\begin{equation}
    \label{eq:int_DD_rep_5th_two}
        I_5 = \frac{63\pi a^8}{32} \int_{0}^{b}  \frac{\rho h \left( 128 h^8 + 2304 h^6 \left( 2r\rho\right)^2 + 6048 h^4 \left( 2r\rho\right)^4 + 3360 h^2 \left( 2r\rho\right)^6 + 315 \left( 2r\rho\right)^8 \right) }{ \left( h^2-\left( 2r\rho\right)^2\right)^{19/2}} \text{d}\rho~.
\end{equation}

The integrated disk-disk repulsion can be written as
\begin{equation}
    U_\text{DD}^\text{R}(r) =  \frac{\pi a^2}{5} A_2 \lambda_a \lambda_b  \left( I_1 + I_2 +I_3 + I_4 +I_5\right)~.
\end{equation}
Its detailed integral form reads
\begin{eqnarray}
    U_\text{DD}^\text{R}(r) &=&  \frac{\pi^2 a^2}{5} A_2 \lambda_a \lambda_b  \int_{0}^{b} \frac{2\rho}{\left( \left(r^2+\rho^2-a^2\right)^2-\left( 2r\rho\right)^2\right)^{19/2}} \left[ 5 \rho^{26}+\left(85 r^2-5 b^2\right) \rho^{24} \right. \nonumber \\
    & & +\left(-120 b^4+800 r^2 b^2-360 r^4\right) \rho^{22} +\left(500 b^6-3880 r^2 b^4+3680 r^4 b^2-280 r^6\right) \rho^{20} \nonumber \\
    & & +\left(-749 b^8+600 r^2 b^6+18900 r^4 b^4-21600 r^6 b^2+3475 r^8\right) \rho^{18} \nonumber \\
    & & +\left(81 b^{10}+19891 r^2 b^8-87600 r^4 b^6+56100 r^6 b^4+16525 r^8 b^2-6525 r^{10}\right) \rho^{16} \nonumber \\
    & & +\left(1296 b^{12}-37376 r^2 b^{10}+64036 r^4 b^8+162400 r^6 b^6-241100 r^8 b^4+49600 r^{10} b^2+3600 r^{12}\right) \rho^{14} \nonumber \\
    & & +\left(-1944 b^{14}+20944 r^2 b^{12}+119364 r^4 b^{10}-509484 r^6 b^8+293100 r^8 b^6+170100 r^{10} b^4 \right. \nonumber \\
    & & \left. \qquad -98000 r^{12} b^2 +3600 r^{14}\right) \rho^{12} \nonumber \\
    & & +\left(1251 b^{16}+10576 r^2 b^{14}-202524 r^4 b^{12}+281376 r^6 b^{10}+435126 r^8 b^8-738000 r^{10} b^6 \right. \nonumber \\
    & & \left. \qquad +170100 r^{12} b^4 +49600 r^{14} b^2-6525 r^{16}\right) \rho^{10} \nonumber \\
    & & +\left(-251 b^{18}-18125 r^2 b^{16}+77500 r^4 b^{14}+202500 r^6 b^{12}-768750 r^8 b^{10}+435126 r^{10} b^8 \right. \nonumber \\
    & & \left. \qquad +293100 r^{12} b^6 -241100 r^{14} b^4+16525 r^{16} b^2+3475 r^{18}\right) \rho^8 \nonumber \\
    & & +\left(-136 b^{20}+6624 r^2 b^{18}+22500 r^4 b^{16}-200000 r^6 b^{14}+202500 r^8 b^{12}+281376 r^{10} b^{10} \right. \nonumber \\
    & & \left. \qquad -509484 r^{12} b^8 +162400 r^{14} b^6+56100 r^{16} b^4-21600 r^{18} b^2-280 r^{20}\right) \rho^6 \nonumber \\
    & & +\left(84 b^{22}+296 r^2 b^{20}-15876 r^4 b^{18}+22500 r^6 b^{16}+77500 r^8 b^{14}-202524 r^{10} b^{12}+119364 r^{12} b^{10} \right. \nonumber \\
    & & \left. \qquad +64036 r^{14} b^8-87600 r^{16} b^6+18900 r^{18} b^4+3680 r^{20} b^2-360 r^{22}\right) \rho^4 \nonumber \\
    & & +\left(-11 b^{24}-424 r^2 b^{22}+296 r^4 b^{20}+6624 r^6 b^{18}-18125 r^8 b^{16}+10576 r^{10} b^{14}+20944 r^{12} b^{12} \right. \nonumber \\
    & & \left. \qquad -37376 r^{14} b^{10}+19891 r^{16} b^8+600 r^{18} b^6-3880 r^{20} b^4+800 r^{22} b^2+85 r^{24}\right) \rho^2 \nonumber \\
    & & + \left( -b^{26}-11 r^2 b^{24}+84 r^4 b^{22}-136 r^6 b^{20}-251 r^8 b^{18}+1251 r^{10} b^{16}-1944 r^{12} b^{14}+1296 r^{14} b^{12} \right. \nonumber \\
    & & \left. \left. \qquad +81 r^{16} b^{10}-749 r^{18} b^8+500 r^{20} b^6-120 r^{22} b^4-5 r^{24} b^2+5 r^{26} \right) \right] \text{d}\rho 
\end{eqnarray}
Completing the integration, we arrive at the final expression of $U_\text{DD}^\text{R}(r)$, which reads
\begin{eqnarray}
    \label{eq:int_DD_rep_final}
    U_\text{DD}^\text{R}(r) &=& \frac{\pi^2 a^2 b^2}{5} A_2 \lambda_a \lambda_b \frac{1}{\left[\left(r+a+b\right)\left(r+a-b\right)\left(r-a+b\right)\left(r-a-b\right)\right]^{17/2}} \left[ 5 r^{22} + 5 r^{20} \left(a^2+b^2\right) \right. \nonumber \\
    & & -5 r^{18} \left(23 a^4-67 a^2 b^2+23 b^4\right)+5 r^{16} \left(a^2+b^2\right) \left(53 a^4-143 a^2 b^2+53 b^4\right) \nonumber \\
    & & -4 r^{14} \left(26 a^8+515 a^6 b^2-1350 a^4 b^4+515 a^2 b^6+26 b^8\right) \nonumber \\
    & & -4 r^{12} \left(a^2+b^2\right) \left(98 a^8-1365 a^6 b^2+2780 a^4 b^4-1365 a^2 b^6+98 b^8\right) \nonumber \\
    & & +2 r^{10} \left(308 a^{12}-1386 a^{10} b^2-4482 a^8 b^4+12165 a^6 b^6-4482 a^4 b^8-1386 a^2 b^{10}+308 b^{12}\right) \nonumber \\
    & & -10 r^8 \left(a^2+b^2\right) \left(32 a^{12}+150 a^{10} b^2-1698 a^8 b^4+3081 a^6 b^6-1698 a^4 b^8+150 a^2 b^{10}+32 b^{12}\right)\nonumber \\
    & & -5 r^6 \left(a^2-b^2\right)^2 \left(a^{12}-438 a^{10} b^2+3 a^8 b^4+2436 a^6 b^6+3 a^4 b^8-438 a^2 b^{10}+b^{12}\right) \nonumber \\
    & & +r^4 \left(a^2-b^2\right)^4 \left(a^2+b^2\right) \left(59 a^8-228 a^6 b^2-2602 a^4 b^4-228 a^2 b^6+59 b^8\right) \nonumber \\
    & & -r^2 \left(a^2-b^2\right)^6 \left(13 a^8+181 a^6 b^2+396 a^4 b^4+181 a^2 b^6+13 b^8\right) \nonumber \\
    & & \left. -\left(a^2-b^2\right)^8 \left(a^2+b^2\right) \left(a^4+5 a^2 b^2+b^4\right) \right]
\end{eqnarray}
As expected, this expression is fully symmetric between the two disks.

In the limit of $a\rightarrow 0$ and $b \rightarrow 0$ under the constraints of $\lambda_a \pi a^2=1$ and $\lambda_b \pi b^2=1$, each disk is reduced to a mass point and the integrated repulsion is reduced to
\begin{equation}
    \lim_{a\rightarrow 0,~b\rightarrow 0} U_\text{DD}^\text{R}(r) = \frac{A_2}{r^{12}}~,
\end{equation}
which is the LJ repulsion between point particles, as expected.

For two disks of the same size ($a=b$), the integrated repulsion is simplified as
\begin{eqnarray}
    \label{eq:int_DD_rep_same_rad}
    U_\text{DD}^\text{R}(r)|_{a=b} &=& \frac{\pi^2 a^4}{5} A_2 \lambda_a^2 \frac{1}{r^9\left(r^2-4a^2\right)^{17/2}} \left( 5 r^{14} +10 r^{12}a^2 
    + 105 r^{10}a^4 -370 r^8 a^6 +1072 r^6 a^8 \right. \nonumber \\ 
    & & \left. -1968 r^4 a^{10} + 2090 r^2 a^{12} -980 a^{14} \right)
\end{eqnarray}

The expression of the repulsive force between two disks can be computed from Eq.~(\ref{eq:int_DD_rep_final}) as
\begin{eqnarray}
    F_\text{DD}^\text{R}(r) &=& -\frac{\partial U_\text{DD}^\text{R}}{\partial r} \nonumber \\
    &=& 12 \pi^2 a^2 b^2 A_2 \lambda_a \lambda_b \frac{r}{\left[\left(r+a+b\right)\left(r+a-b\right)\left(r-a+b\right)\left(r-a-b\right)\right]^{19/2}} \left[ r^{24} + 2 r^{22} \left(a^2+b^2\right) \right. \nonumber \\
    & & -2 r^{20} \left(16 a^4-47 a^2 b^2+16 b^4\right) \nonumber \\
    & & +2 r^{18} \left(a^2+b^2\right) \left(37 a^4-100 a^2 b^2+37 b^4\right) \nonumber \\
    & & -r^{16} \left(9 a^8+850 a^6 b^2-2100 a^4 b^4+850 a^2 b^6+9 b^8\right) \nonumber \\
    & & -4 r^{14} \left(a^2+b^2\right) \left(51 a^8-635 a^6 b^2+1285 a^4 b^4-635 a^2 b^6+51 b^8\right) \nonumber \\
    & & +4 r^{12} \left(84 a^{12}-364 a^{10} b^2-1421 a^8 b^4+3775 a^6 b^6-1421 a^4 b^8-364 a^2 b^{10}+84 b^{12}\right) \nonumber \\
    & & -4 r^{10} \left(a^2+b^2\right) \left(51 a^{12}+313 a^{10} b^2-3484 a^8 b^4+6415 a^6 b^6-3484 a^4 b^8+313 a^2 b^{10}+51 b^{12}\right) \nonumber \\
    & & -r^8 \left(9 a^{16}-2336 a^{14} b^2+5684 a^{12} b^4+11724 a^{10} b^6-30750 a^8 b^8 \right. \nonumber \\
    & & \left. \qquad +11724 a^6 b^{10}+5684 a^4 b^{12}-2336 a^2 b^{14}+9 b^{16}\right) \nonumber \\
    & & +2 r^6 \left(a^2-b^2\right)^2 \left(a^2+b^2\right) \left(37 a^{12}-388 a^{10} b^2-1651 a^8 b^4+5474 a^6 b^6-1651 a^4 b^8-388 a^2 b^{10}+37 b^{12}\right) \nonumber \\
    & & -2 r^4 \left(a^2-b^2\right)^4 \left(16 a^{12}+127 a^{10} b^2-638 a^8 b^4-1950 a^6 b^6-638 a^4 b^8+127 a^2 b^{10}+16 b^{12}\right) \nonumber \\
    & & +2 r^2 \left(a^2-b^2\right)^6 \left(a^2+b^2\right) \left(a^8+52 a^6 b^2+188 a^4 b^4+52 a^2 b^6+b^8\right) \nonumber \\
    & & \left. +\left(a^2-b^2\right)^8 \left(a^8+10 a^6 b^2+20 a^4 b^4+10 a^2 b^6+b^8\right)  \right]~.
\end{eqnarray}
%This force is radial and positive definite, as expected for repulsion.

The repulsive force between two disks with the same radius (i.e., $a=b$) reads
\begin{eqnarray}
    F_\text{DD}^\text{R}(r)|_{a=b} &=& 12 \pi^2 a^4 A_2 \lambda_a^2 \frac{1}{r^{10}\left(r^2-4a^2\right)^{19/2}} \left( r^{16}+4 r^{14} a^2 +30 r^{12} a^4 -104 r^{10} a^6 +382 r^8 a^8 -936 r^6 a^{10} \right. \nonumber \\
    & & \left. \qquad \qquad \qquad \qquad \qquad \qquad \qquad +1492 r^4 a^{12} -1400 r^2 a^{14} +588 a^{16} \right)~.
\end{eqnarray}
This expression can also be obtained by taking the derivative of the corresponding potential in Eq.~(\ref{eq:int_DD_rep_same_rad}) with respect to $r$.

\section{Wall Potentials for Disks and LJ Point Particles in 2D}

\subsection{Integrated Point-Wall Potential}

\begin{figure}[htbp]
    \centering
    \includegraphics[width=0.5\textwidth]{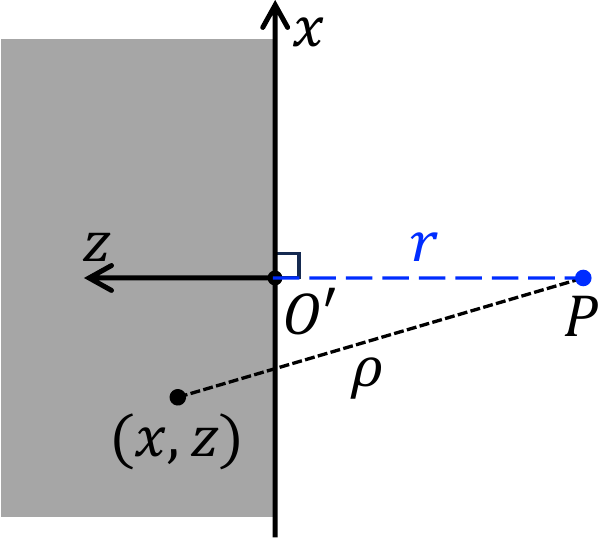}
    \caption{A Cartesian coordinate system describing a 2D configuration of a point mass located at a distance $r$ from the boundary of a half space.}
    \label{fg:point-wall-geo}
\end{figure}

The integrated point-wall potential in 2D can be obtained by integrating the LJ potential between a point particle and a half-space. Using the coordinate system in Fig.~\ref{fg:point-wall-geo}, the integrated potential can be written as
\begin{eqnarray}
    \label{eq:int_PW}
    U_\text{PW}(r) &=& \lambda_w \int_{-\infty}^{\infty} \int_{0}^{\infty}  U_\text{LJ}(\rho)\vert_{\rho=\sqrt{\left( z+r \right)^2 + x^2}} \text{d}x \text{d}z \nonumber \\
    &=& \lambda_w \int_{-\infty}^{\infty} \int_{0}^{\infty}  \left[ \frac{A_2}{\left( \left( z+r \right)^2 + x^2 \right)^6} - \frac{A_1}{\left( \left( z+r \right)^2 + x^2 \right)^3} \right] \text{d}x \text{d}z \nonumber \\
    &=& \frac{3 \pi \lambda_w}{32} \left( \frac{21}{80} \frac{A_2}{r^{10}} - \frac{A_1}{r^4} \right)~,
\end{eqnarray}
where $\lambda_w$ is the areal density of LJ point masses making up the half-space and $r$ is the distance of the point particle ($P$ in Fig.~\ref{fg:point-wall-geo}) from the boundary of the half-space (i.e., the location of the wall). Here $\lambda_w$ is assumed to be constant. The corresponding force expression is
\begin{eqnarray}
    \label{eq:force_PW}
    F_\text{PW}(r) = \frac{3 \pi \lambda_w}{8} \left( \frac{21}{32} \frac{A_2}{r^{11}} - \frac{A_1}{r^5} \right)~.
\end{eqnarray}

\subsection{Integrated Disk-Wall Potential}

\begin{figure}[htb]
    \centering
    \includegraphics[width=0.5\textwidth]{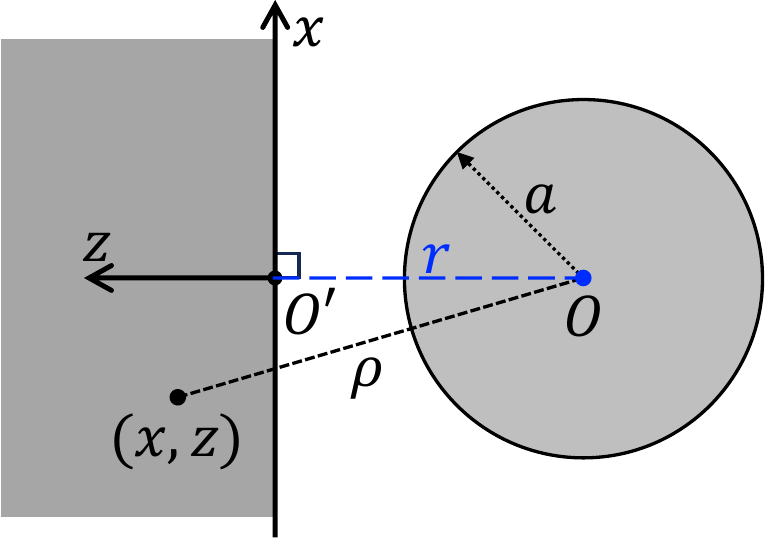}
    \caption{A Cartesian coordinate system describing a 2D configuration of a disk of radius $a$ with its center located at a distance $r$ from the boundary of a half space.}
    \label{fg:disk-wall-geo}
\end{figure}

The integrated disk-wall potential can be obtained by integrating the disk-point potential presented in the previous section over a half-space consisting of uniformly distributed LJ material points. Using the coordinate system in Fig.~\ref{fg:disk-wall-geo}, the integrated disk-wall potential is
\begin{equation}
    \label{eq:int_DW}
    U_\text{DW}(r) = \lambda_w \int_{-\infty}^{\infty} \int_{0}^{\infty}  U_\text{DP}(\rho)\vert_{\rho=\sqrt{\left( z+r \right)^2 + x^2}}~ \text{d}x \text{d}z~.
\end{equation}
Similar to $U_\text{DP}$, the integrated disk-wall potential $U_\text{DW}(r)$ can also be split into attractive and repulsive terms. The attractive terms is
\begin{eqnarray}
    \label{eq:int_DW_attr}
    U_\text{DW}^\text{A}(r) &=& \lambda_w \int_{-\infty}^{\infty} \int_{0}^{\infty}  U_\text{DP}^\text{A}(\rho)\vert_{\rho=\sqrt{\left( z+r \right)^2 + x^2}} \text{d}x \text{d}z \nonumber \\
    &=& -\frac{\pi}{2} A_1 \lambda_w \lambda_a a^2 \int_{-\infty}^{\infty} \int_{0}^{\infty} \frac{2 \left( z+r \right)^2 + 2 x^2 + a^2}{\left( \left( z+r \right)^2 + x^2 - a^2 \right)^4} \text{d}x \text{d}z \nonumber \\
    &=& -\frac{3 \pi^2}{32} A_1 \lambda_w \lambda_a a^2 \frac{r}{\left( r^2 - a^2 \right)^{5/2}}~.
\end{eqnarray}
The repulsive term is
\begin{eqnarray}
    \label{eq:int_DW_rep}
    U_\text{DW}^\text{R}(r) &=& \lambda_w \int_{-\infty}^{\infty} \int_{0}^{\infty}  U_\text{DP}^\text{R}(\rho)\vert_{\rho=\sqrt{\left( z+r \right)^2 + x^2}} \text{d}x \text{d}z \nonumber \\
    &=& \frac{\pi}{5} A_2 \lambda_w \lambda_a a^2 \int_{-\infty}^{\infty} \int_{0}^{\infty} \nonumber \\
    & & \frac{5 \left( \left( z+r \right)^2 + x^2 \right)^4 +40 a^2 \left( \left( z+r \right)^2 + x^2 \right)^3 +60 a^4 \left( \left( z+r \right)^2 + x^2 \right)^2 +20 a^6 \left( \left( z+r \right)^2 + x^2 \right) + a^8}{\left( \left( z+r \right)^2 + x^2 - a^2 \right)^{10}} \text{d}x \text{d}z \nonumber \\
    &=& \frac{63 \pi^2}{163840} A_2 \lambda_w \lambda_a a^2 \frac{\left(64 r^6 + 336 r^4 a^2 + 280 r^2 a^4 + 35 a^6 \right) r}{\left( r^2 - a^2 \right)^{17/2}}~.
\end{eqnarray}

With the expression of $U_\text{DW}^\text{A}(r)$ and $U_\text{DW}^\text{R}(r)$, the attractive and repulsive forces between the disk and the wall (i.e., the half-space) can be calculated by taking the derivative with respect to $r$. The results are as follows.
\begin{equation}
    F_\text{DW}^\text{A} (r) = -\frac{3 \pi^2}{32} A_1 \lambda_w \lambda_a a^2 \frac{4r^2+a^2}{\left( r^2 - a^2 \right)^{7/2}}~,
\end{equation}
and
\begin{equation}
    F_\text{DW}^\text{R} = \frac{63 \pi^2}{32768} A_2 \lambda_w \lambda_a a^2 \frac{ 128 r^8 +896 a^2 r^6 +1120 a^4 r^4 +280 a^6 r^2 + 7 a^8 }{\left( r^2 - a^2 \right)^{19/2}}
\end{equation}
%It is clear that the attractive force is negative while the repulsive force is positive, consistent with the convention.

In the limit of $a \rightarrow 0$ under the constraint of $\lambda_a \pi a^2 = 1$, the integrated disk-wall potential is reduced to the point-wall potential, as expected.

\section{Verification of Analytical Results}

\begin{figure}[htb]
    \centering
    \includegraphics[width=1.0\textwidth]{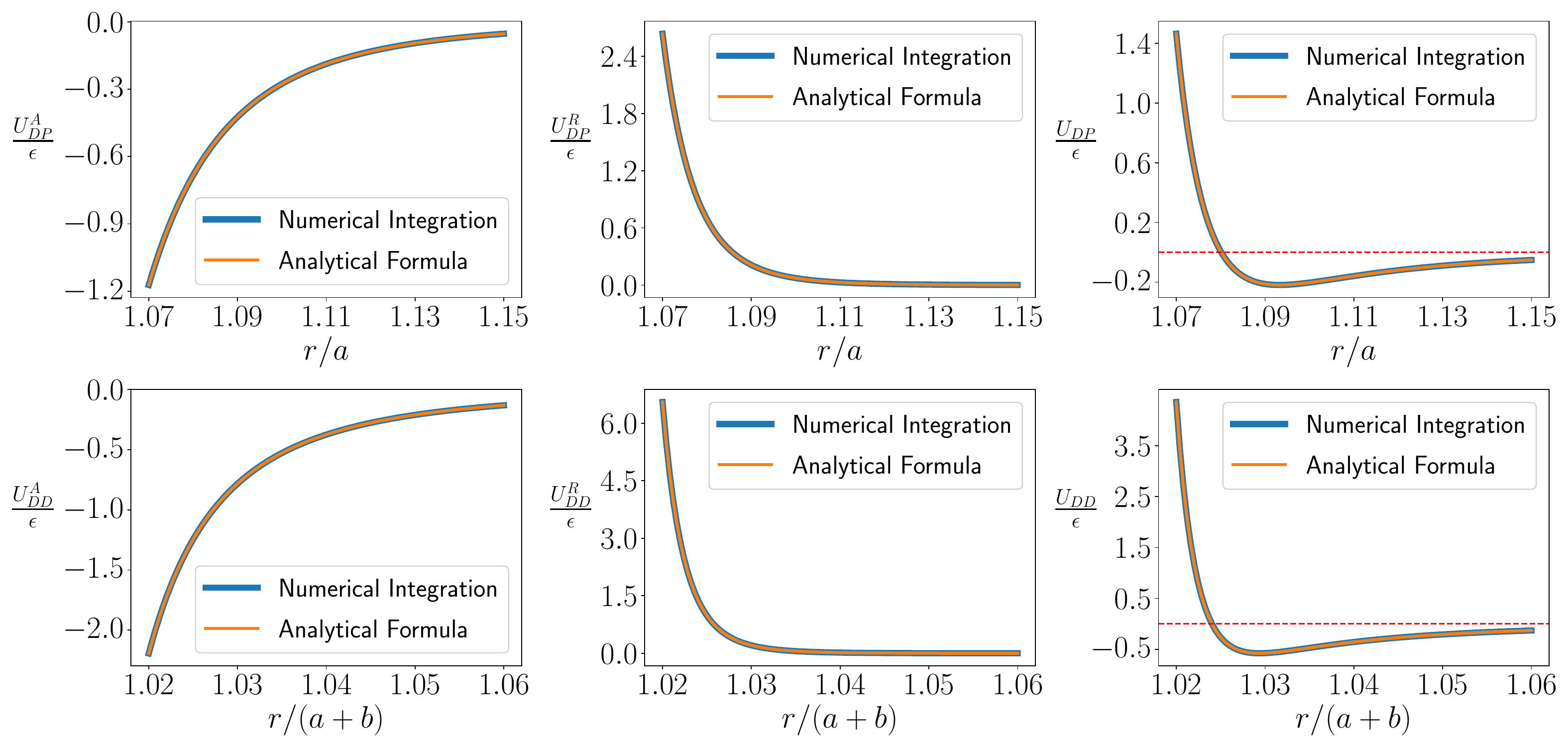}
    \caption{Comparison between analytical results and numerical integration for the disk-point (first row) and disk-disk (second row) potentials: attraction (left), repulsion (middle), and full potential (right). For the disk-point potential, the radius of the disk is $10\sigma$. For the disk-disk potential, the radii of the two disks are $10\sigma$ and $15\sigma$, respectively.}
    \label{fg:DP_DD_verify}
\end{figure}

The analytical results on the disk-point and disk-disk potentials are further verified by comparing them with the results from direct numerical integration of the Lennard-Jones potential. The method of numerical integration is described in detail elsewhere.\cite{Wang2025EPJE} The comparisons included in Figs.~\ref{fg:DP_DD_verify} show perfect agreement between the two and completely verify the analytical results presented here.

\begin{figure}[htb]
    \centering
    \includegraphics[width=0.4\textwidth]{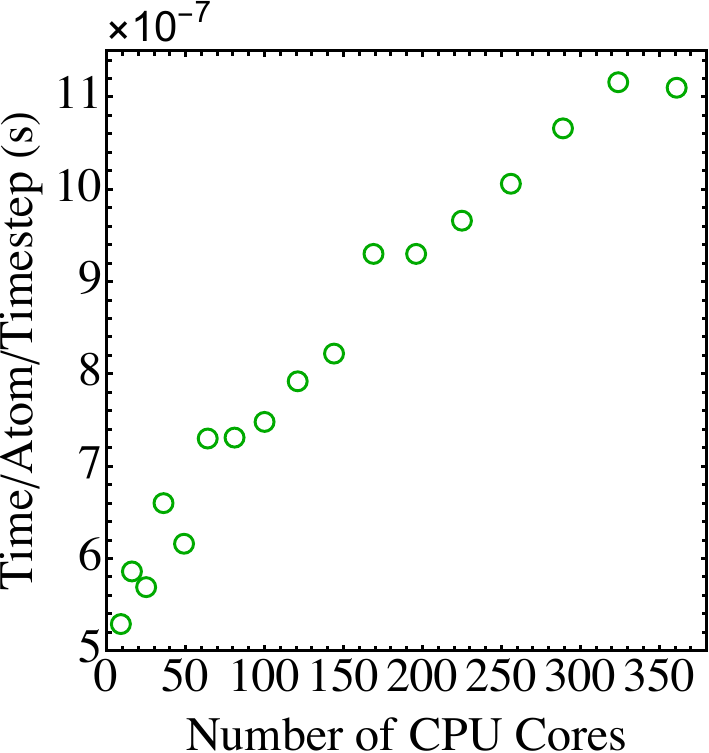}
    \caption{Wall time per particle per timestep is plotted against the number of CPU cores used for the simulation. The benchmark system contains $401,155$ particles.}
    \label{fg:performance}
\end{figure}

% implementation of analytical potentials in LAMMPS
\section{Implementation in LAMMPS}

The integrated disk-point, disk-disk, point-wall, and disk-wall potentials presented in the previous sections were implemented in the LAMMPS molecular dynamics simulator as a new \texttt{pair\_style} and \texttt{fix wall}.\cite{LAMMPS} The \texttt{pair\_style disk} was developed on the basis of the existing \texttt{pair\_style colloid} by replacing the integrated LJ potentials between spheres by those between disks.\cite{LAMMPS_COLLOID} The code iterates over all point particles and disks in a neighbor list of a given point particle or disk and classifies each pair based on their interaction type (i.e., point-point, disk-point, or disk-disk). Then it computes the corresponding interaction energy and force using the appropriate integrated potential. A new \texttt{fix wall/disk} is also implemented for the interaction between a point particle or a disk and a wall (i.e., the boundary of a half-space in two dimensions). All these new commands are grouped into a new user package \texttt{DISK}. {\color{blue}The package is available for download on Github:xxx}. This new package enables a LAMMPS user to simulate two-dimensional systems of disks, which may include an explicit solvent modeled as a Lennard-Jones liquid.

To assess the parallel efficiency of the disk potentials implemented in LAMMPS, we performed a series of benchmark simulations. The test system consisted of $400,899$ LJ solvent atoms and $256$ disks with a diameter of $20\sigma$. The total number of particles is therefore $401,155$. All benchmarks were executed on three computational nodes, each equipped with an AMD EPYC 7702 128-core CPU and 256 GB of RAM. To ensure an optimal domain decomposition for the 2D simulation box, we selected core counts that are perfect squares: 9 ($3\times 3$), 16 ($4\times 4$), 25 ($5\times 5$), 36 ($6\times 6$), 49 ($7\times 7$), 64 ($8\times 8$), 81 ($9\times 9$), 100 ($10\times 10$), and 121 ($11\times 11$). All simulations were performed using the 2-Apr-2025 version of LAMMPS for $500,000$ timesteps with \texttt{neighbor multi} enabled, which were known to dramatically enhance the performance of LAMMPS to simulate colloid systems with large size ratios. The wall time per particle per timestep as a function of the number of CPU cores used is shown in Fig.~\ref{fg:performance}. This time was approximately $5\times 10^{-7}$ seconds for a small number of CPU cores and increased by about a factor of $2$ to $\sim 1\times 10^{-6}$ seconds when more than 300 cores were used and the number of particles on each core is less than $1,500$. In terms of efficiency, the DISK package implemented in LAMMPS outperforms the existing COLLOID package, which is designed for three-dimensional colloidal systems. This trend is expected as each particle in two dimensions has a fewer number of neighbors than in three dimensions. Furthermore, the performance of the DISK package can be further improved in the future through code optimization.

\section{Liquid-Vapor Coexistence of the Lennard-Jones Fluid in Two Dimensions}

\begin{figure}[htb]
    \centering
    \includegraphics[width=1.0\textwidth]{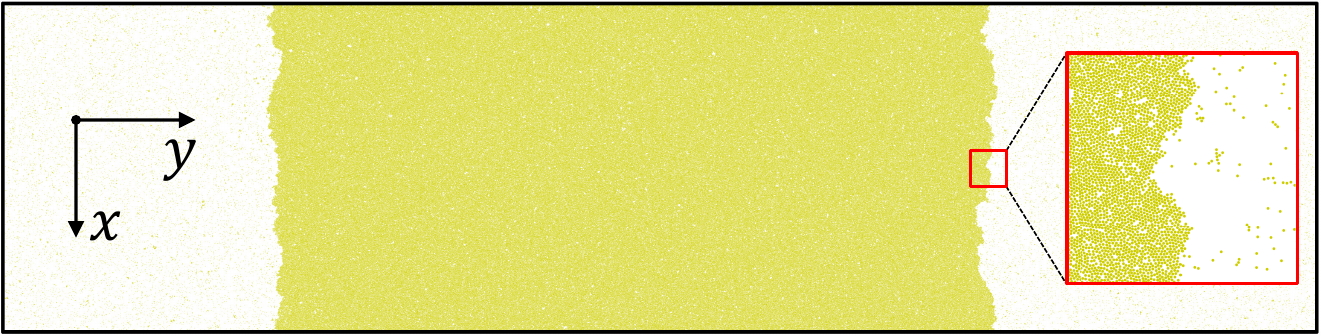}
    \caption{Visualization of the liquid-vapor coexistence of the Lennard-Jones fluid at temperature $T=0.42 \epsilon/k_\text{B}$ in two dimensions. The roughness of each liquid-vapor interface is visible, especially in the magnified visualization shown in the inset.}
    \label{fg:lv_coexist}
\end{figure}

The binodal (i.e., the coexistence curve) of the Lennard-Jones fluid, consisting of point particles interacting with each other through the Lennard-Jones 12-6 potential with a cutoff of $3.0\sigma$, is obtained by examining the liquid and vapor densities at coexistence. To this end, $10^6$ point particles are placed in a simulation box of $800\sigma$ along the $x$-axis and $3200\sigma$ along the $y$-axis, with periodic boundary conditions applied to all directions. At a target temperature ($T$), every particle is coupled to a thermal bath at that temperature through a Langevin thermostat with a damping rate of $0.01 \tau^{-1}$. The value of $T$ ranges from $0.4\epsilon/k_\text{B}$ to $0.5\epsilon/k_\text{B}$. After equilibration, a liquid film emerges in the simulation box with its two edges fluctuating along the $x$-axis. One visualization of the system at $T=0.42 \epsilon/k_\text{B}$ is shown in Fig.~\ref{fg:lv_coexist}.

\section{Thermalization Schemes for Evaporating Suspensions}

The simulation methods are explained in detail in the main text. Here, we use a binary disk suspension to illustrate the methodology of modeling a drying process. The disk suspension is placed in a simulation box of $800\sigma$ along the $x$-axis and $1600\sigma$ along the $y$-axis, with periodic boundary conditions applied along the $x$ direction. The entire system is confined by two walls in the $y$ direction, one lower wall located at $y=0$ and a higher wall located at $y=1600\sigma$. The interactions with the walls are governed by the point-wall and disk-wall potentials derived in the previous sections. The suspension has an initial thickness of about $800\sigma$ and occupies the half simulation box closer to the lower wall, which is attractive to the solvent with a $3\sigma$ cutoff for the corresponding point-wall potential. The solvent vapor fills the other half of the simulation box and the higher wall is purely repulsive for the solvent with the corresponding point-wall potential truncated at its minimum. The adsorption of the solvent on the top wall is thus prevented. The system is well equilibrated prior to drying. To evaporate solvent out of the disk suspension, the vapor in the deletion zone, a slab within $100\sigma$ from the higher wall, is removed at a given rate, which sets the solvent evaporation rate.

\begin{figure}[htb]
    \centering
    \includegraphics[width=1.0\textwidth]{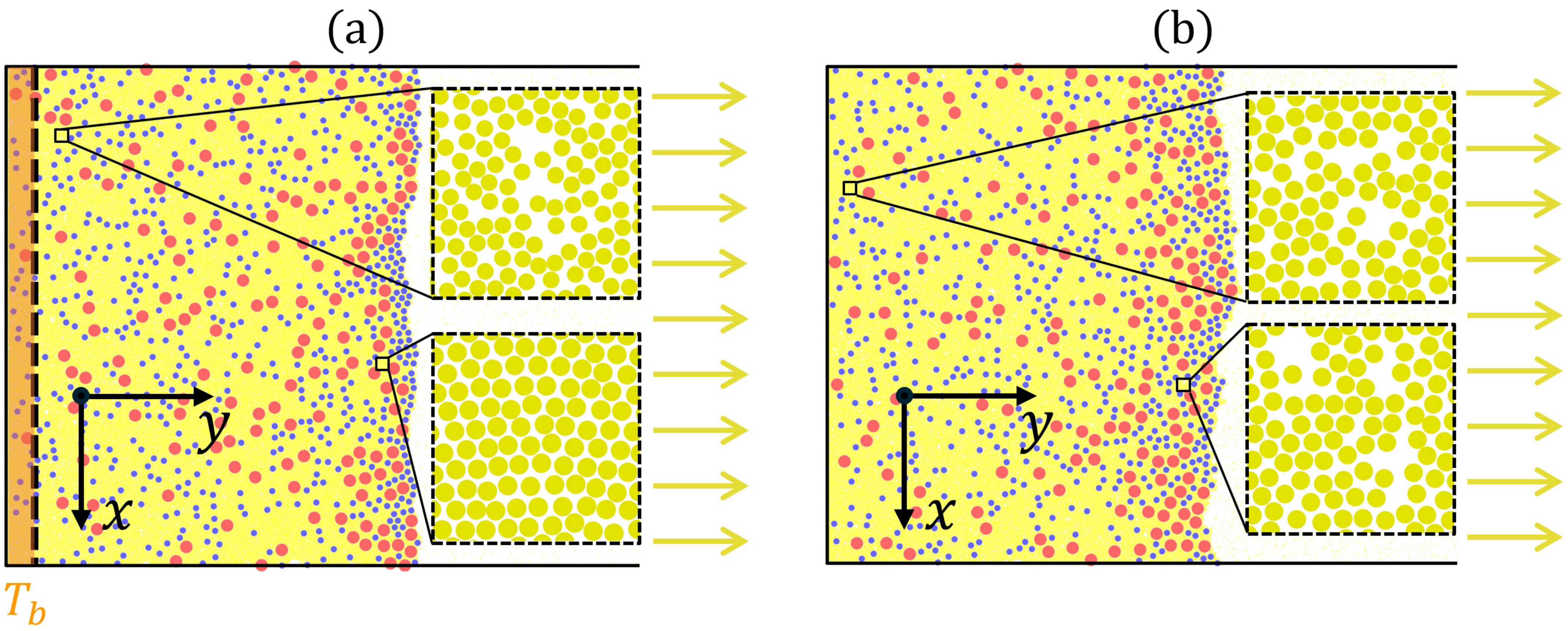}
    \caption{Visualization of evaporating binary disk suspensions in 2D. (a) Langevin thermostat was used to thermalize the solvent particles in a thin layer, which is adjacent to the lower wall located at the left boundary (i.e., $y=0$) of the simulation box, at temperature $T_b=0.42\epsilon/k_\text{B}$. Because of evaporative cooling, temperature in the suspension decreases from $T_b$ in the thermalized layer to a much lower value at the surface of the suspension (see Fig.~\ref{fg:temperature_profile} below). As a result, the solvent near the suspension surface solidifies, as shown in the inset. (b) All the solvent particles in the simulation box are thermalized at temperature $T_b$ with a DPD thermostat. As a result, the solvent in the suspension maintains its liquid state during drying.}
    \label{fg:langevin_dpd}
\end{figure}

\begin{figure}[htb]
    \centering
    \includegraphics[width=0.5\textwidth]{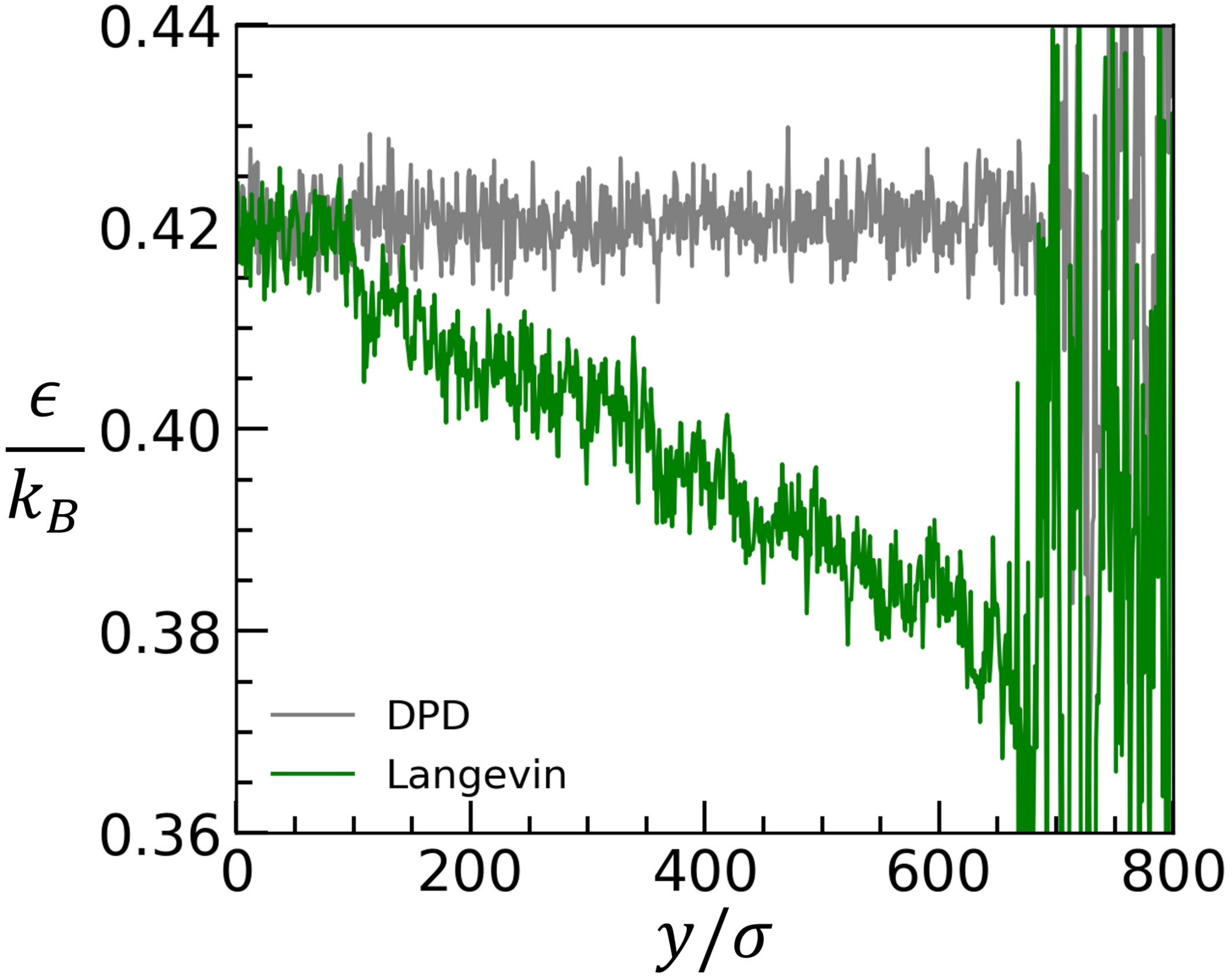}
    \caption{Comparison of temperature profiles along the $y$-axis (i.e., the direction of drying) from simulations using the Langevin thermostat (green) and those using the DPD thermostat (grey).}
    \label{fg:temperature_profile}
\end{figure}

An important issue in modeling solvent evaporation is how the system is thermalized. A frequently adopted strategy is to thermalize a thin layer of solvent adjacent to the lower wall at a constant temperature ($T_b$) with a weak Langevin thermostat,\cite{Cheng2011JCP} as shown in Fig.~\ref{fg:langevin_dpd}(a) where $T_b = 0.42\epsilon/k_\text{B}$. All of the disks and other solvent particles move according to Newtonian mechanics. Due to evaporative cooling, temperature decreases from $T_b$ to a lower value toward the liquid-vapor interface where evaporation occurs. A temperature profile along the $y$ direction when about 25\% of the solvent have evaporated is shown in Fig.~\ref{fg:temperature_profile}. The temperature at the evaporating interface decreases to about $0.38\epsilon/k_\text{B}$. At this temperature, the solvent near the interface freezes and transitions to a solid state, as shown in Fig.~\ref{fg:langevin_dpd}(a) by directly visualizing the solvent particles in a local region. Therefore, such a thermalization scheme is appropriate for simulating the drying process of a disk suspension. We turn to a different strategy where all the solvent particles in the simulation box are thermalized with a thermostat based on dissipative particle dynamics (DPD), which conserves momentum locally and is suitable for simulating a solvent evaporation process.\cite{Liu2023SM}

The DPD thermostat is based on the following set of equations.
\begin{eqnarray}
    \label{eq:dpd}
    & & \vec{f} = (F^D + F^R) \hat{r}_{ij}, \qquad r < r_c \\
    & & F^D = -\gamma w^2(r) (\hat{r}_{ij} \cdot \vec{v}_{ij}) \\
    & & F^R = \alpha\, \sqrt{2k_\text{B}T\gamma/\delta t}\, w(r) \\
    & & w(r) = 1 - r/r_c
\end{eqnarray}
Here, $\vec{f}$ is the friction added to particle $i$ when its separation ($r$) from particle $j$ is less than the cutoff $r_c$. The friction applied to particle $j$ is therefore $-\vec{f}$. $\hat{r}_{ij}$ is the unit vector that points from particle $j$ to particle $i$. The friction contains two terms, a dissipative force $F^D$ and a random force $F^R$. $\gamma$ is a coefficient. $w(r)$ is a weighting factor defined in the last equation above. $\vec{v}_{ij}$ is the velocity of particle $i$ with respect to that of particle $j$. $\alpha$ is a Gaussian random number with zero mean and unit variance. $k_\text{B}$ is the Boltzmann constant. $T$ is the target temperature. $\delta t$ is the MD timestep. For the simulations reported here, $\gamma = 1.0 \sqrt{m \epsilon/\sigma^2}$ and $ T=0.42 \epsilon/k_\text{B}$.

The DPD thermostat is applied to all solvent particles in the simulation box. After the suspension is equilibrated, solvent evaporation is initiated and solvent vapor is removed out of the simulation box at a given rate. All solvent particles remaining in the simulation box are still subjected to the DPD thermostat. The temperature profile in Fig.~\ref{fg:temperature_profile} shows that the DPD thermalization scheme ensures that the entire simulation box is maintained at a target temperature and evaporative cooling is avoided. A careful examination of the solvent state, as shown in Fig.~\ref{fg:langevin_dpd}(b), confirms that the solvent remains liquid throughout the suspension during evaporation (i.e., vapor removal). All of the evaporation simulations reported in the main text are conducted with this DPD thermalization scheme.

\end{document}